\documentclass[pdflatex,sn-mathphys-num]{sn-jnl}

\usepackage{graphicx}%
\usepackage{multirow}%
\usepackage{geometry}
\usepackage{amsmath,amssymb,amsfonts}%
\usepackage{amsthm}%
\usepackage{mathrsfs}%
\usepackage[title]{appendix}%
\usepackage{xcolor}%
\usepackage{textcomp}%
\usepackage{manyfoot}%
\usepackage{booktabs}%
\usepackage{algorithm}%
\usepackage{algorithmicx}%
\usepackage{algpseudocode}%
\usepackage{listings}%
\usepackage{booktabs, multirow, amsmath, amssymb, xcolor, graphicx, cleveref, enumitem}

\theoremstyle{thmstyleone}%
\theoremstyle{thmstyletwo}%

\theoremstyle{thmstylethree}%

\begin{document}

\title[Article Title]{Bridging the Post-discharge Gap: A Traceable Multi-agent Framework for Safe and Continuous Care}


\author[1]{\fnm{Runwei} \sur{Guan}}\email{runwayrwguan@hkust-gz.edu.cn}
\equalcont{These authors contributed equally to this work.}

\author[1]{\fnm{Yi} \sur{Zhou}}\email{yz@hkust-gz.edu.cn}
\equalcont{These authors contributed equally to this work.}

\author[1]{\fnm{Heyi} \sur{Lin}}\email{hlinaw@connect.hkust-gz.edu.cn}
\equalcont{These authors contributed equally to this work.}

\author[1]{\fnm{Jinjing} \sur{Zhu}}\email{jinjingzhu.mail@gmail.com}
\equalcont{These authors contributed equally to this work.}

\author[2]{\fnm{Mingyuan} \sur{Hou}}\email{amoshou@distinctclinic.com}
\equalcont{These authors contributed equally to this work.}


\author[1]{\fnm{Yang} \sur{Yang}}\email{yyang945@connect.hkust-gz.edu.cn}

\author[2]{\fnm{Fang} \sur{Yuan}}\email{adayuan@distinctclinic.com}

\author[2]{\fnm{Xiaohong} \sur{Lin}}\email{xiaohonglin@distinctclinic.com}

\author[1]{\fnm{Shaofeng} \sur{Liang}}\email{shawnsfliang@hkust-gz.edu.cn}

\author[1]{\fnm{Xuming} \sur{Hu}}\email{xuminghu@hkust-gz.edu.cn}

\author[2]{\fnm{Tao} \sur{Li}}\email{tao.li@distinctclinic.com}

\author[2]{\fnm{Tianbin} \sur{Zhao}}\email{berniezhao@distinctclinic.com}

\author[1]{\fnm{Yutao} \sur{Yue}}\email{yutaoyue@hkust-gz.edu.cn}

\author[1,2]{\fnm{Zhiyuan} \sur{Wang}}\email{philipwang@distinctclinic.com}

\author*[1]{\fnm{Hui} \sur{Xiong}}\email{xionghui@hkust-gz.edu.cn}

\affil*[1]{\orgdiv{Thrust of Artificial Intelligence}, \orgname{The Hong Kong University of Science and Technology (Guangzhou)}, \orgaddress{\city{Guangzhou}, \postcode{511453}, \state{Guangdong}, \country{China}}}

\affil[2]{\orgname{Distinct Healthcare}, \orgaddress{\city{Shenzhen}, \postcode{10587}, \state{Guangdong}, \country{China}}}

\abstract{
Post-discharge clinical follow-up is critical for maintaining continuity of care and mitigating long-term health risks. However, traditional follow-up paradigms suffer from shortage of health workforce, fragmented patient histories, and information silos across clinical departments. While large language models have demonstrated potential in medical question-answering, their deployment in continuous care is hindered by hallucination risks and a fundamental inability to reason over longitudinal, patient-specific constraints. Here we present Healink, a memory-enhanced multi-agent framework to support AI-assisted post-discharge follow-up by generating prescription-grounded, traceable responses that improved completeness and perceived clinical utility in retrospective and physician-blinded
evaluations. The architecture seamlessly integrates a triage routing mechanism, a unified memory enhancement module utilizing a robust relational database for optimal latency, and a strict constraint-based retrieval-augmented generation engine. By vectorizing historical clinical records and employing weighted similarity functions across diverse phenotypic and intervention dimensions, Healink ensures precise inter-patient and intra-patient case matching while actively preventing cross-departmental drug conflicts. We evaluated Healink on a dataset comprising 400 continuous and 85 highly complex real-world follow-up cases, alongside the webMedQA benchmark. In a rigorous single-blind evaluation conducted by clinical experts, the framework outperformed human physician baselines in both authoritativeness and clinical safety. Crucially, our analysis reveals a notable divergence between automated metric scores and physician preferences, highlighting the indispensability of expert-in-the-loop validation for clinical artificial intelligence. By generating a traceable, white-box evidence chain, Healink provides a scalable, safe, and highly effective paradigm for intelligent patient management, ultimately enhancing societal healthcare outcomes.
}




\maketitle

\section{Introduction}\label{sec1}

The provision of continuous healthcare extends far beyond the initial clinical encounter \cite{cuzick2023importance}. Post-discharge follow-up represents an essential component of clinical practice that significantly reduces hospital readmission rates and optimizes long-term patient prognoses \cite{balasubramanian2025outpatient}. Despite its clinical importance, continuous patient management remains structurally inefficient within modern healthcare systems. The current paradigm relies heavily on manual physician interventions, rendering it highly susceptible to workforce shortages and delayed responses \cite{chen2024digital}. Furthermore, modern healthcare delivery often results in fragmented medical records distributed across disparate specialized departments. This information isolation creates substantial risks for patients with complex comorbidities, frequently leading to unrecognized cross-departmental drug interactions and suboptimal joint-treatment strategies \cite{10.1001/jamainternmed.2021.4878}.

Recent advancements in large language models have catalyzed the development of automated medical dialogue systems \cite{ong2026large,liu2025generalist,liu2023medical}. However, the application of general-purpose language models in diagnostic follow-up is constrained by severe inherent limitations \cite{shool2025systematic}. Conventional models lack the capacity for longitudinal patient memory and frequently generate hallucinatory recommendations when confronted with sparse clinical context \cite{peng2023study}. Previous approaches have attempted to mitigate these issues through distinct architectural strategies, yet each leaves critical gaps for continuous follow-up care. Retrieval-augmented generation systems such as MedRAG \cite{zhao2025medrag} and MedGraph-RAG \cite{wu2025medical} enhance factual grounding through external knowledge retrieval but treat patient queries as isolated instances without structured memory across sessions. Multi-agent frameworks such as TriageMD \cite{liu2026multi} and MedAgent \cite{li2024mmedagent} coordinate multiple reasoning modules yet lack hard-constrained prescription anchoring and cross-department joint decision-making. Multimodal systems such as MIMo \cite{chen2025mimo} demonstrate visual reasoning capabilities but require substantial training infrastructure and do not provide structured patient memory or cross-session coordination. Even recent safety-focused constellations such as Polaris \cite{mukherjee2024polaris} and traceable diagnostic systems such as DeepRare \cite{zhao2026agentic} do not simultaneously satisfy the full set of architectural requirements for safe follow-up care.

To clarify these distinctions, we compare eight representative systems \cite{singhal2025toward,tu2025towards,liu2025multi,tang2024medagents} across five clinically relevant dimensions: whether the system requires domain-specific training (training-free deployment), whether it maintains structured patient profiles across sessions (structured patient memory), whether it enforces prescription-based constraints at the architecture level (hard-constraint prescription anchoring), whether it coordinates care across multiple departments (cross-department joint decision-making), and whether it provides auditable evidence for every clinical claim (white-box evidence chain) (Table \ref{tab:comparison}). This comparison reveals that no existing system simultaneously satisfies all five criteria. The absence of this combination explains why prior systems, despite impressive performance on diagnostic benchmarks, have not achieved deployment in continuous follow-up settings where patient safety depends on longitudinal consistency and pharmacological constraint enforcement.

Moreover, existing medical AI systems typically require substantial computational resources for domain-specific training or fine-tuning \cite{mukherjee2024polaris,jin2023medcpt,zhao2026agentic,palepu2025towards}, creating a deployment barrier that excludes resource-constrained healthcare systems from accessing AI-augmented clinical tools. Healink addresses this inequity by operating as a training-free architecture that leverages off-the-shelf foundation models through constraint-based orchestration rather than parameter modification, enabling zero-overhead deployment across diverse resource settings.

To overcome these structural and algorithmic limitations, we introduce Healink, a specialized multi-agent framework engineered for highly reliable, traceable, and personalized clinical follow-up. The system architecture abandons the conventional monolithic model approach \cite{singhal2025toward,zhang2023huatuogpt,chen2025mimo,wu2025medical,zhao2025medrag,mukherjee2024polaris,zhao2026agentic} in favor of a highly coordinated multi-agent state graph  \cite{liu2026multi}. The workflow initiates with an emergency routing module designed to strictly intercept life-threatening queries and malicious prompts. To optimize system latency and streamline computational overhead, the query rewriting and contextual memory functions are unified into a singular memory agent. This module verifies patient identity alignment, extracts non-incremental physical parameters, and reconstructs the query into a patient-specific clinical formulation. Subsequently, the framework employs global session caching coupled with dynamic summarization to penetrate cross-session contexts, thereby enabling joint decision-making for patients concurrently consulting multiple departments.

A core methodological innovation of Healink lies in its dual-pathway retrieval engine powered by robust relational databases and high-dimensional embeddings. For intra-patient historical matching, the system executes fine-grained semantic retrieval across longitudinal records to establish a personalized disease evolution baseline. Concurrently, an advanced knowledge reasoning engine deconstructs complex clinical intents into atomic queries for coreference resolution. To ensure the clinical validity of inter-patient reference cases, we introduce a hard-constrained filtering mechanism that executes physical interception prior to vector retrieval. Candidate cases must strictly satisfy developmental stage alignments and diagnostic classifications before undergoing multi-dimensional weighted semantic evaluation. The final retrieval score is derived from a weighted combination of symptom phenotypes, intervention strategies, and follow-up feedback metrics, ensuring that only highly actionable, evidence-based data inform the final generation.

\begin{table}
\setlength\tabcolsep{2.8pt}
\centering
\caption{Multi-dimensional comparison of clinical follow-up AI systems. \checkmark = fully supported, - = partial support, \texttimes = not supported.}
\vspace{-4mm}
\label{tab:comparison}
\begin{tabular}{lccccc}
\toprule
\multirow{2}{*}{\textbf{Systems}} & \multirow{2}{*}{\textbf{Training-free}} & \textbf{Structured} & \textbf{Hard-constraint} & \textbf{Cross-dept} & \textbf{White-box} \\
 & & \textbf{patient memory} & \textbf{prescription} & \textbf{joint decision} & \textbf{evidence chain} \\
\midrule
Med-PaLM 2 \cite{singhal2025toward} & \texttimes & \texttimes & \texttimes & \texttimes & \texttimes \\
AMIE \cite{tu2025towards} & \texttimes & \texttimes & \texttimes & \texttimes & \texttimes \\
TriageMD \cite{liu2025multi} & \checkmark & - & \texttimes & \texttimes & \texttimes \\
MedAgent \cite{tang2024medagents} & \texttimes & - & \texttimes & \texttimes & \texttimes \\
MIMo \cite{chen2025mimo} & \texttimes & \texttimes & \texttimes & \texttimes & - \\
MedGraph-RAG \cite{wu2025medical} & \texttimes & \texttimes & \texttimes & \texttimes & \checkmark \\
MedRAG \cite{zhao2025medrag} & \checkmark & \texttimes & \texttimes & \texttimes & - \\
Polaris \cite{mukherjee2024polaris} & \texttimes & - & - & \texttimes & - \\
DeepRare \cite{zhao2026agentic} & \checkmark & - & - & \texttimes & \checkmark \\
Healink (Ours) & \checkmark & \checkmark & \checkmark & \checkmark & \checkmark \\
\bottomrule
\end{tabular}
\end{table}

We comprehensively evaluated Healink using 400 standard and 85 complex real-world clinical follow-up scenarios. The system generated a fully transparent, white-box evidence chain that maps directly to authoritative medical guidelines and literature. Under single-blind peer review by practicing physicians, Healink consistently surpassed the clinical quality and safety thresholds of human doctors. Importantly, our empirical observations highlight a critical phenomenon in medical artificial intelligence evaluation. We identified a distinct divergence between the performance metrics assigned by generalized language models and the qualitative preferences of clinical practitioners. While automated judges prioritize linguistic fluency and broad comprehensive coverage, human physicians place an absolute premium on clinical safety, conciseness, and actionable logic. This discrepancy validates our multi-agent constraint design and underscores the necessity of anchoring medical artificial intelligence evaluation in expert clinical judgment rather than relying solely on computational benchmarks.

\begin{figure}
    \centering
    \includegraphics[width=0.99\linewidth]{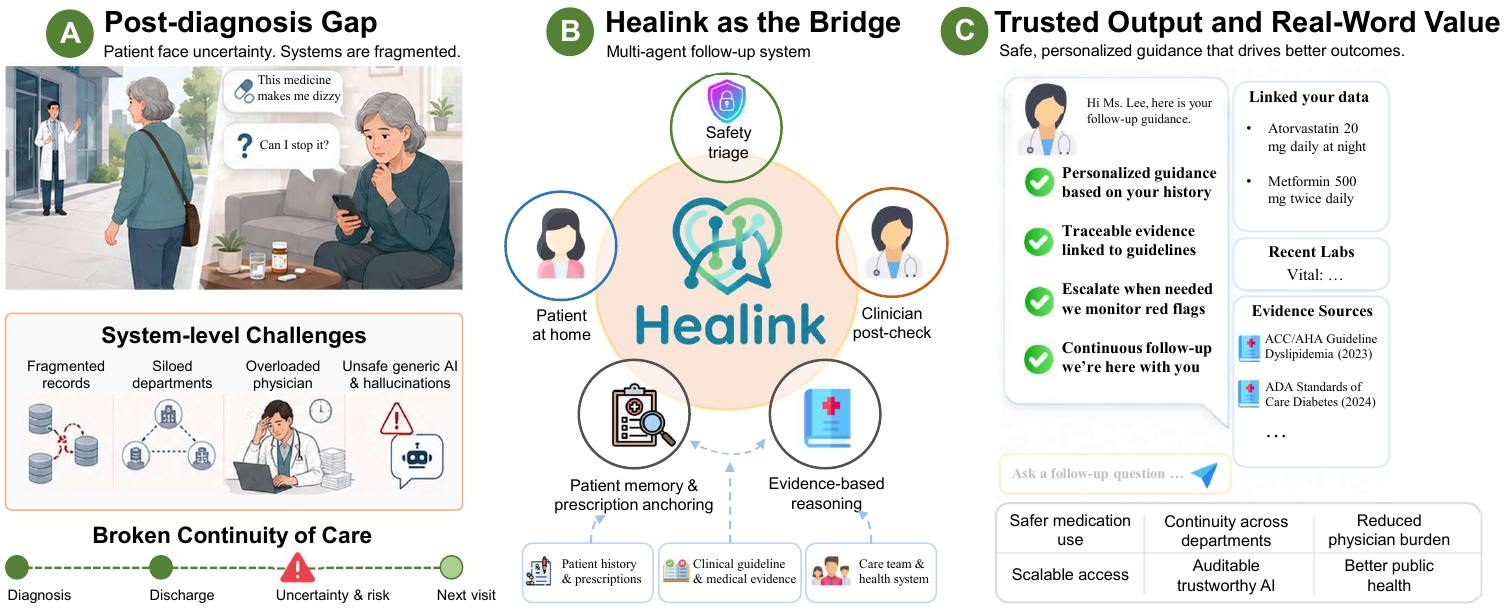}
    \caption{Healink bridges the post-discharge gap through a multi-agent follow-up system. (A) Patient uncertainty and systemic fragmentation (disconnected records, siloed departments, physician overload, unsafe AI) create broken continuity of care. (B) Healink's multi-agent architecture integrates safety triage, patient-specific memory anchoring, evidence-based reasoning, and clinician validation. (C) The system delivers personalized, guideline-grounded guidance with traceable evidence, improving medication safety, care continuity, and clinical efficiency.}
    \label{fig:placeholder}
\end{figure}

\section{Results}
\label{sec:results}

\subsection{Evaluation on real-world follow-up data}

\begin{figure}
    \centering
    \includegraphics[width=0.99\linewidth]{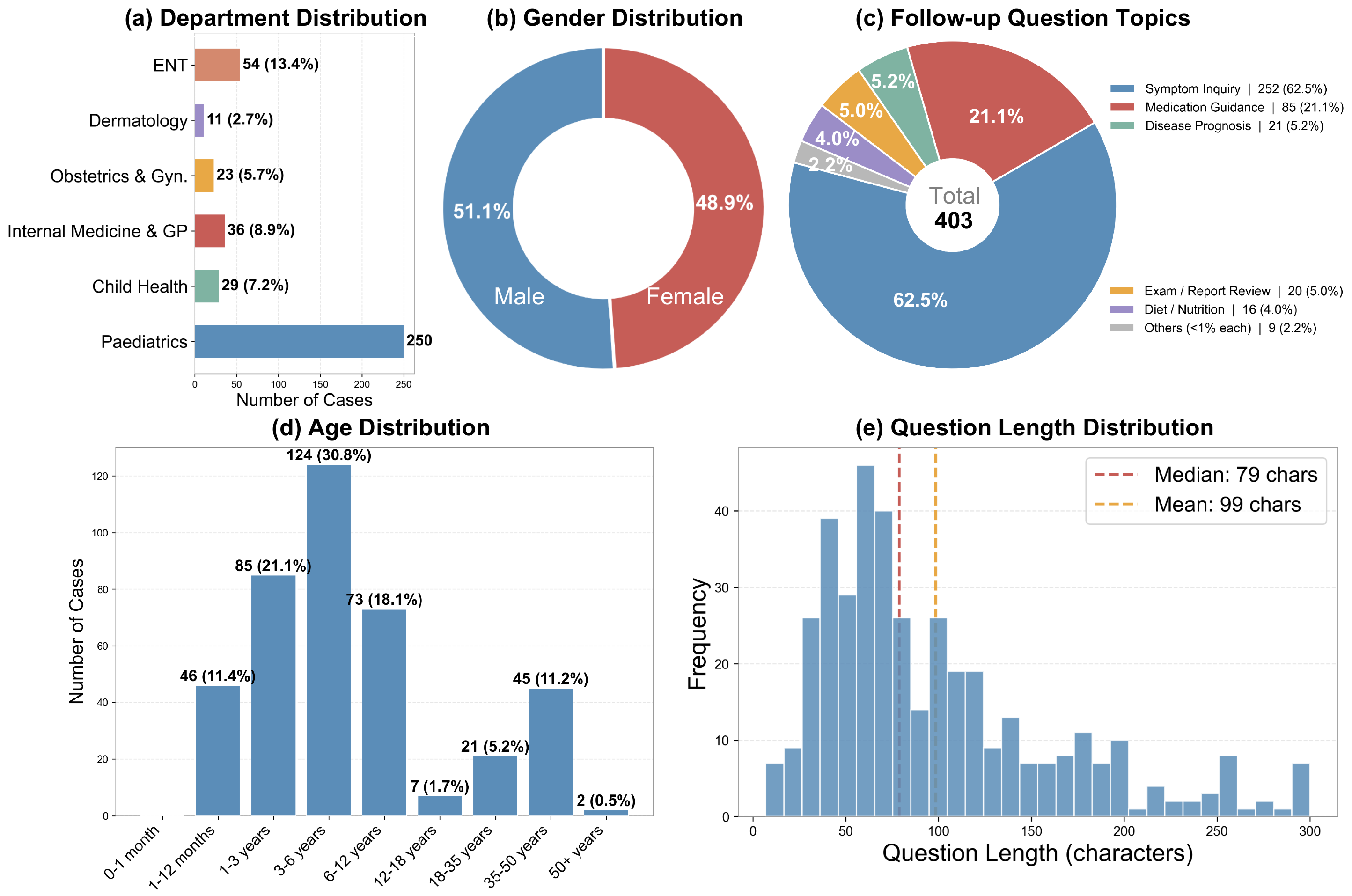}
    \caption{Dataset characteristics of the 400-case real-world follow-up evaluation set. (a) Department distribution across six specialties. (b) Gender distribution. (c) Follow-up question topic distribution classified by semantics. (d) Age distribution across nine age groups. (e) Question length distribution showing median and mean values.}
    \label{fig:400_case_distribution}
\end{figure}

\begin{figure}
    \centering
    \includegraphics[width=0.99\linewidth]{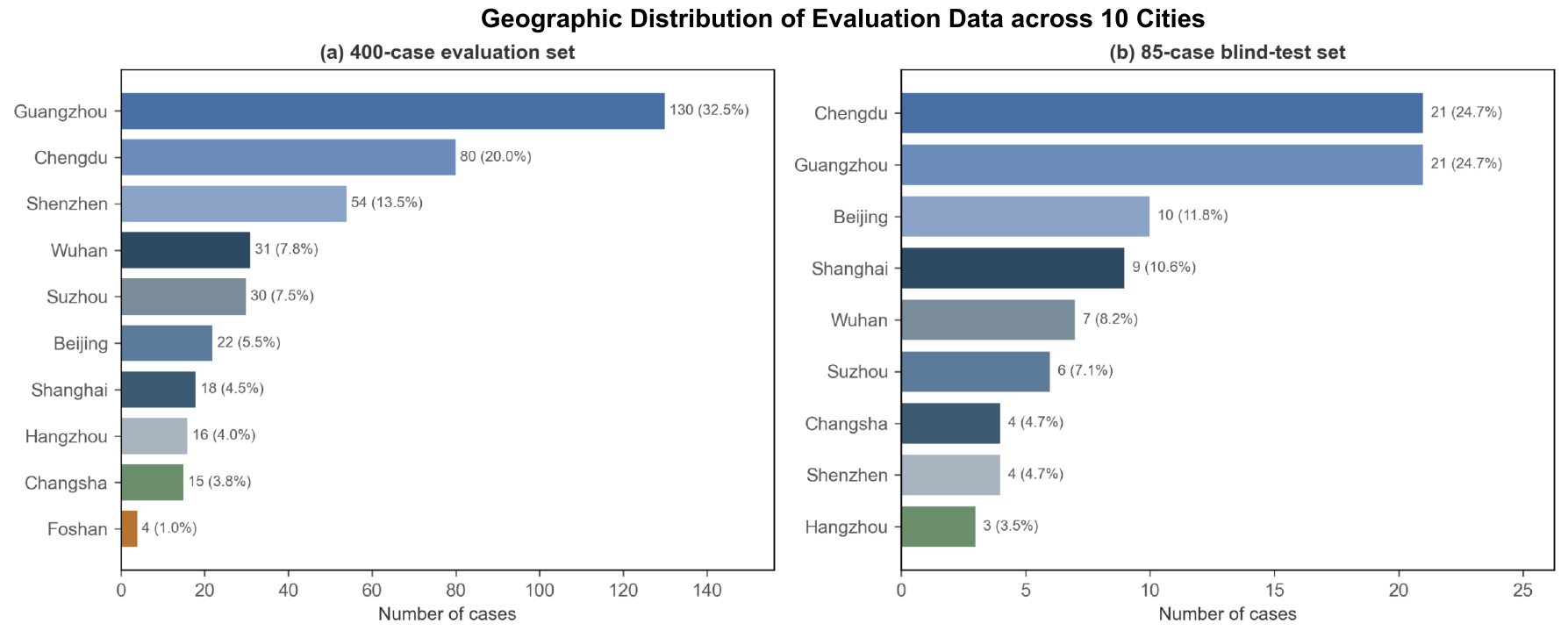}
    \caption{The geographical distribution of the two real-world follow-up evaluation datasets from clinics and hospitals.}
    \label{fig:geographics}
\end{figure}

\subsubsection{Dataset construction and rationale}

\begin{figure}
    \centering
    \includegraphics[width=0.99\linewidth]{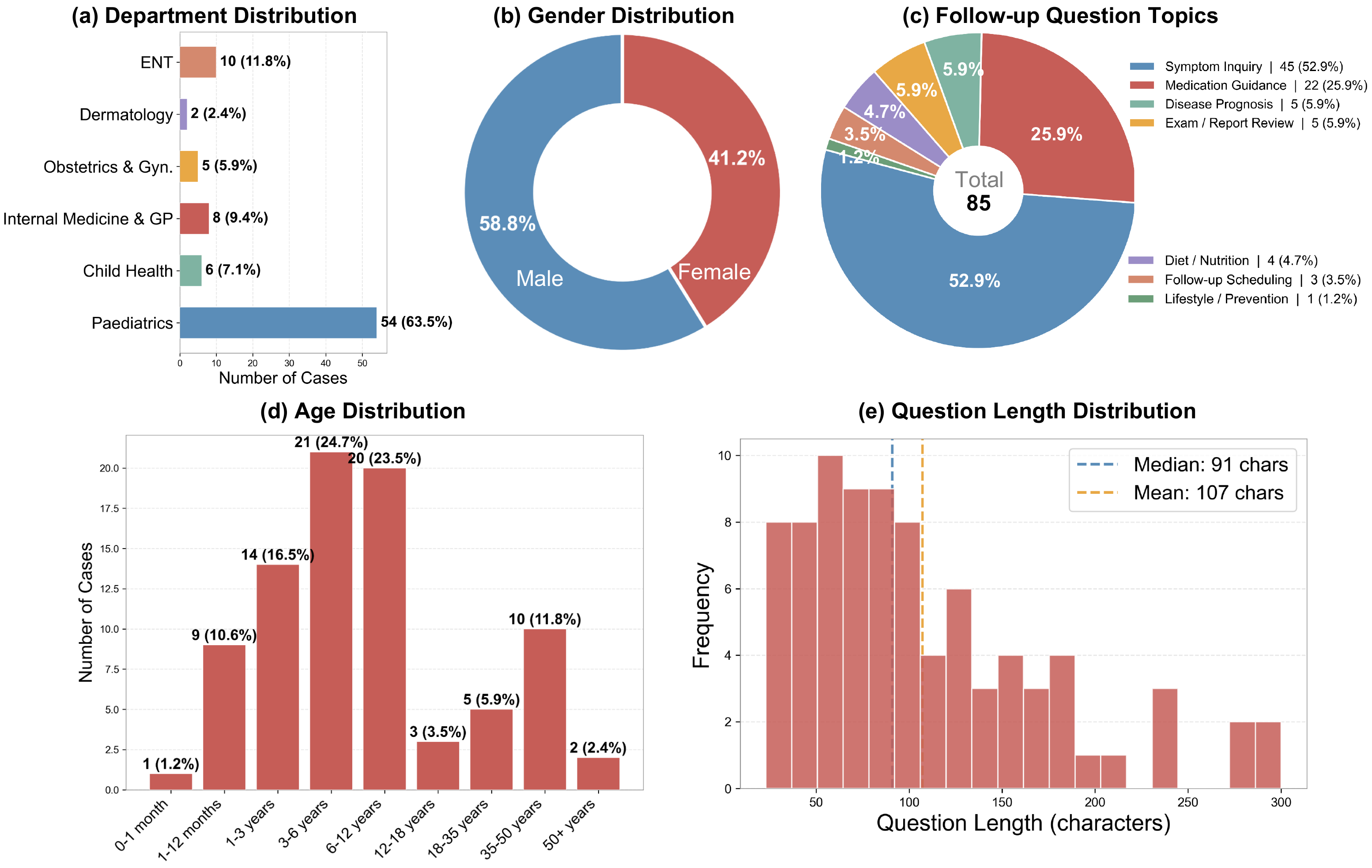}
    \caption{Characteristics of the 85-case blind-test evaluation set. (a) Department distribution across six specialties. (b) Gender distribution. (c) Follow-up question topic distribution. (d) Age group distribution. (e) Question length distribution with median and mean indicated.}
    \label{fig:85_case_distribution}
\end{figure}

A critical challenge in evaluating follow-up AI systems is the absence of established benchmarks that capture the unique characteristics of post-discharge interactions. Unlike diagnostic datasets, where the gold standard is a definitive diagnosis, follow-up clinical references are inherently pluralistic: a medication query, a symptom report, and a scheduling request each demand different information structures. To address this gap, we constructed a two-tier evaluation corpus comprising 400 real-world follow-up question-and-answer pairs and a physician-blinded subset of 85 cases. The detailed statistics of two evaluation datasets are shown in Fig. \ref{fig:400_case_distribution} and Fig. \ref{fig:85_case_distribution}. Moreover, as Fig. \ref{fig:geographics} shows, the 400-case and 85-case evaluation sets were drawn from 28 clinical sites spanning 11 major cities across China, including Beijing, Shanghai, Guangzhou, Shenzhen, Chengdu, and Hangzhou, ensuring broad demographic and geographic representation. This multi-center sampling strategy was deliberately designed to capture diverse patient populations, regional practice variations, and cross-specialty clinical contexts, thereby enhancing the generalizability and ecological validity of our findings.

The 400-case dataset was drawn from a 5-month window (June-October 2025) across six specialties at a tertiary multi-specialty clinic: Paediatrics, Child Health, Internal Medicine and General Practice, Obstetrics and Gynaecology, Dermatology, and Otorhinolaryngology (ENT). These six specialties were selected to span the three dominant categories of post-discharge inquiry: medication management (Paediatrics, Internal Medicine), condition monitoring (Child Health, ENT), and procedure-related follow-up (Obstetrics and Gynaecology, Dermatology). Importantly, all cases were first-round, text-only interactions where the patient had an established diagnosis and active prescription, directly mirroring the operational context for which Healink was designed.

It is important to clarify the nature of the physician-authored responses used as evaluation references.  These responses were generated by practicing physicians under real-world workflow constraints (managing 40 to 60 follow-up inquiries per shift), and thus reflect clinical practice quality rather than an idealized medical optimum.  We use the term clinical practice reference rather than "gold standard" to acknowledge this distinction.  While these responses represent the actual standard of care that patients currently receive, they are not presumed to be medically exhaustive.  This framing is methodologically intentional: it allows us to evaluate whether AI systems can meet or exceed the current standard of clinical practice, rather than comparing against an aspirational ideal that may not reflect realistic care delivery.  The Expert Consensus Reference, discussed below, provides a complementary benchmark.

Each case comprised three components: the patient's original query (with demographic and clinical metadata), the corresponding physician-authored response serving as the clinical practice reference, and parallel AI-generated responses from Healink and nine baseline configurations. Two complementary evaluation methodologies were employed.

\textbf{Automated LLM-based evaluation (all 400 cases).} All 400 cases were scored on Authoritativeness and Clinical Reference Consistency (CRC) using structured rubrics (see Table \ref{tab:auth_rubric} and Table \ref{tab:gold_rubric} in Section \ref{sec:eval_criteria}) by GPT-4-turbo (OpenAI) \cite{achiam2023gpt}, a model chosen specifically to be \emph{independent} from all backend LLMs used in Healink and the baseline systems. This independence is methodologically essential: using the same model family as either the system under test or the baselines would introduce evaluator bias, whether through stylistic preference or knowledge-base overlap. GPT-4-turbo was prompted to perform point-by-point identification and matching of key information between each AI-generated response and the physician-authored clinical reference, producing structured JSON outputs containing scored decompositions of every response. Authoritativeness (1-10) assessed alignment with mainstream evidence-based medical consensus. Gold Standard Consistency (1-10) measured the information coverage ratio between AI-generated and physician-authored responses, computed as:

\begin{equation}
\label{eq:gold_standard}
S_{\text{CRC}} = \min\left(10, \max\left(1, \left\lfloor \frac{N_{\text{covered}}}{N_{\text{total}}} \times 10 + 0.5 \right\rfloor \right)\right)
\end{equation}
where $N_{\text{covered}}$ is the number of gold-standard information points captured by the AI response (meaning-identical or semantically equivalent), $N_{\text{total}}$ is the total number of information points identified in the gold standard, and $\lfloor \cdot \rceil$ denotes rounding to the nearest integer. If $N_{\text{total}} = 0$ (indicating a non-informative gold standard), $S_{\text{CRC}}$ defaults to 10. Explicit penalties are applied for factually contradictory or unsafe recommendations (see Table \ref{tab:gold_rubric}).

\begin{table}
\centering
\caption{Random sampling statistics of 85-case single-blind evaluation.}
\vspace{-4mm}
\label{tab:follow-up-stats}
\setlength\tabcolsep{3.0pt}
\begin{tabular}{lccccc}
\toprule
\textbf{Department} & \textbf{Total Follow-ups} & \textbf{Text-only Follow-ups} & \textbf{Sampling Rate} & \textbf{Sample Size} & \textbf{Doctor Num} \\
\midrule
Pediatrics & 13259 & 4912 & 1.1\% & 54 & 9 \\
Child Healthcare & 3378 & 519 & 1.2\% & 6 & 1 \\
ENT & 2410 & 964 & 1.0\% & 10 & 2 \\
Internal Medicine & 1880 & 728 & 1.1\% & 8 & 2 \\
Dermatology & 945 & 203 & 1.0\% & 2 & 1 \\
Gynecology & 1566 & 446 & 1.1\% & 5 & 1 \\
\midrule
\textbf{Total} & \textbf{25308} & \textbf{8562} & \textbf{1.0\%} & \textbf{85} & \textbf{16} \\
\bottomrule
\end{tabular}
\end{table}

\textbf{Physician-blinded evaluation (85-case subset).} A stratified subset of 85 cases underwent physician-blinded review by 16 physicians across all six specialties, who scored each response on Accuracy (0-3) and Completeness (0-3) without knowledge of the generating system (see Table \ref{tab:blind_rubric} in Section \ref{sec:eval_criteria}). Stratification ensured proportional specialty representation, with the subset selected to maximize clinical diversity while controlling for question-type distribution.

This dual-tier design serves distinct methodological purposes. The 400-case evaluation provides statistical power for quantitative comparison across model configurations, while the blinded subset tests a subtler hypothesis: whether physicians, when stripped of authorship information, can distinguish AI-generated follow-up from human-authored responses, and which they judge to be clinically superior. The finding that blinded physicians consistently rated AI responses above their own peers' (Section \ref{sec:blinded}) carries implications that extend beyond Healink to the broader question of how follow-up care quality is defined and measured.

\subsubsection{Main results on authoritativeness and clinical reference consistency}

\begin{figure}
    \centering
    \includegraphics[width=0.99\linewidth]{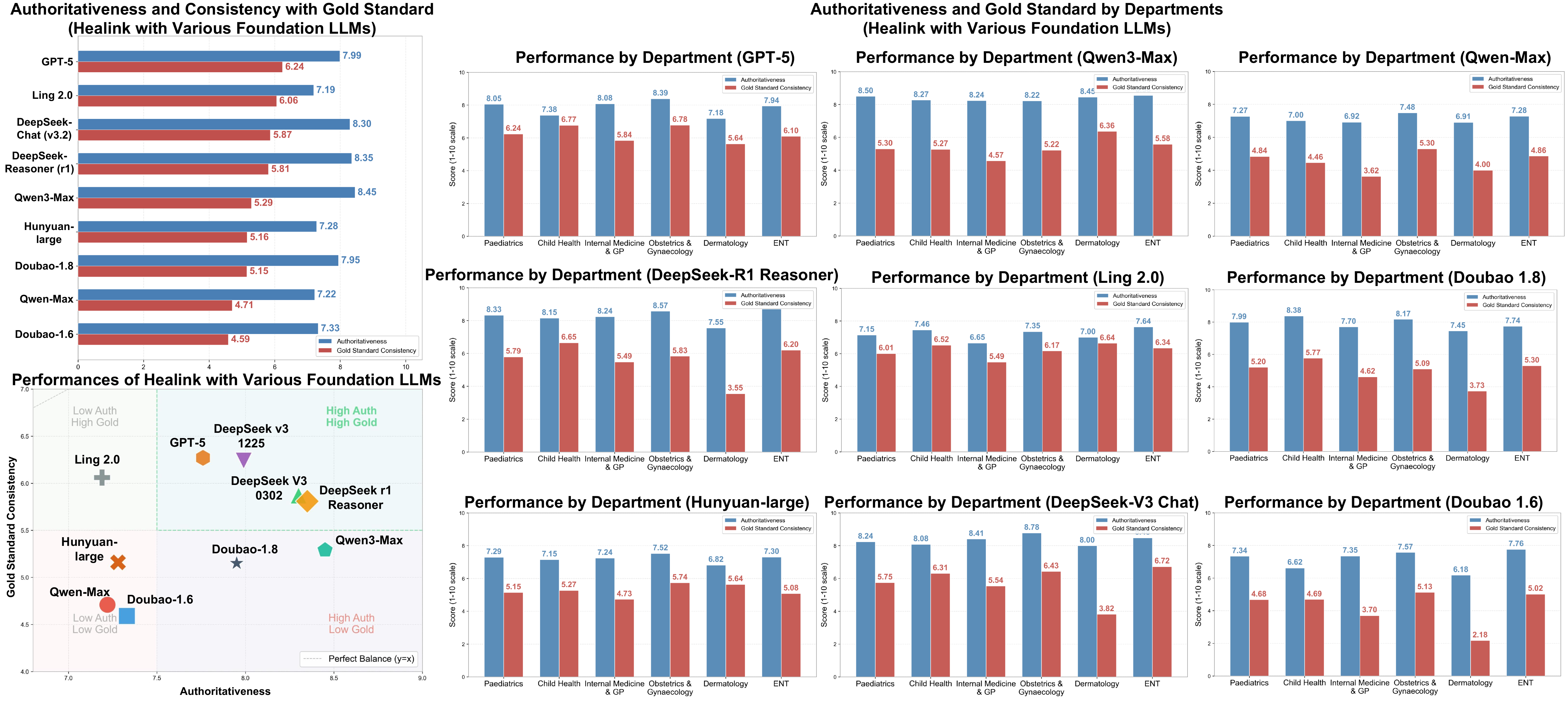}
    \caption{Performance of Healink with various foundation LLMs on the 400-case dataset. Left panels: Authoritativeness and Consistency with Clinical Reference scores by department for each backend LLM, including GPT-5 \cite{singh2025openai}, Qwen series \cite{qwen3}, DeepSeek series \cite{guo2025deepseek}, Ant Ling 2.0 \cite{li2025every}, Doubao (SEED) series  \cite{seed2025seed} and Hunyuan-large \cite{sun2024hunyuan}. Right panel: overall comparison across all model configurations.}
    \label{fig:400_case_eval}
\end{figure}

\begin{figure}
    \centering
    \includegraphics[width=0.99\linewidth]{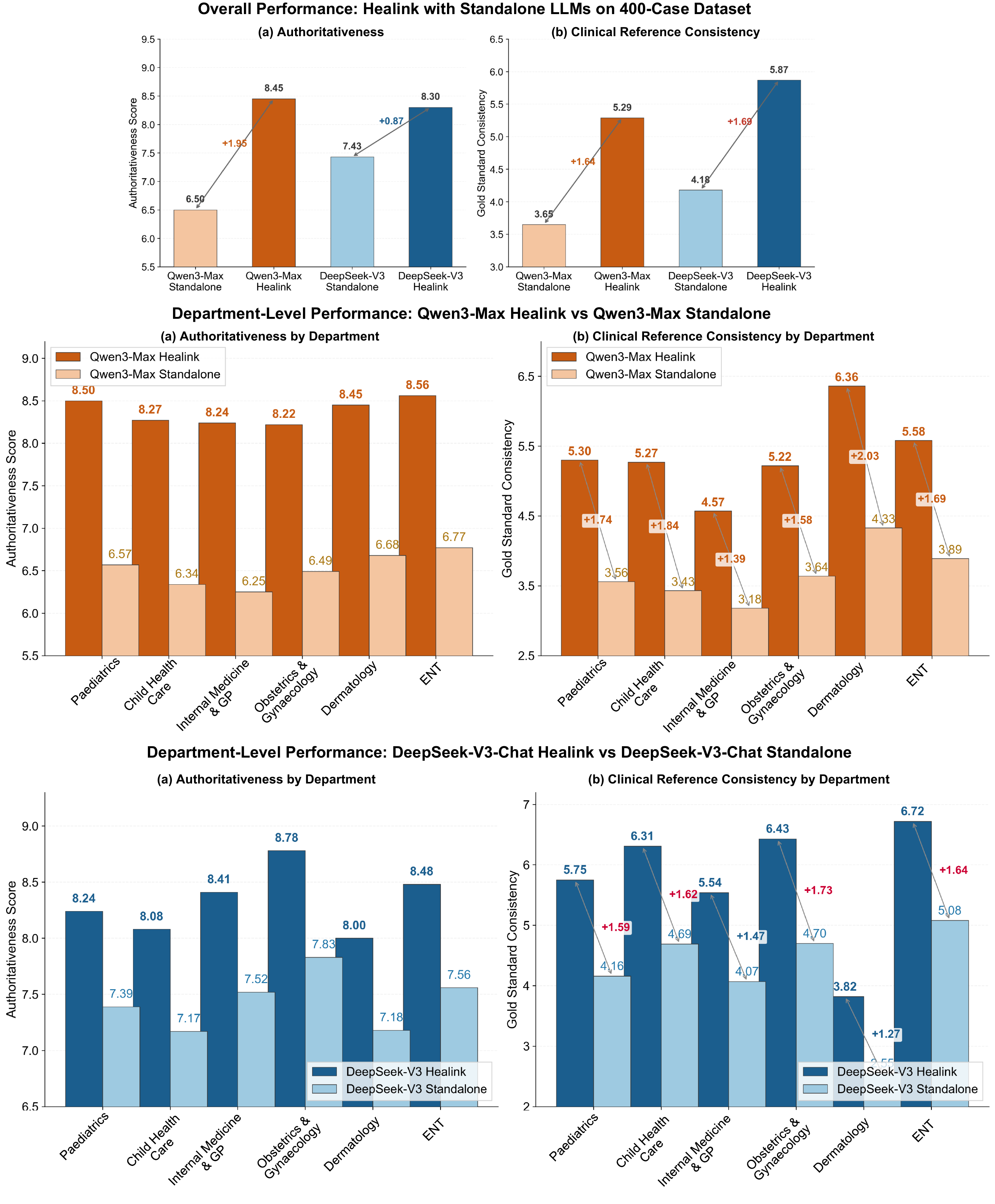}
    \vspace{-4mm}
    \caption{Comparison of performances on the 400-case dataset by Healink and standalone LLMs. Two metrics of authoritativeness and clinical reference consistency are included.}
    \label{fig:healink_vs_single_model}
\end{figure}

\textbf{Healink achieves consistent performance across nine foundation LLMs.} Healink is a multi-agent framework that orchestrates specialized functional modules atop an underlying foundation LLM. Because the architecture is designed to be model-agnostic, we first assessed how Healink performs when paired with a diverse set of foundation LLMs, ranging from frontier commercial APIs to cost-effective open-source alternatives. Fig.~\ref{fig:400_case_eval} presents the overall and department-level performance of Healink with nine foundation LLMs on the 400-case evaluation set.

\textbf{Information completeness varies more than factual authority across backends.} On Authoritativeness, Healink with DeepSeek-Chat v3.2 achieved the highest score (8.41), closely followed by Qwen3-Max (8.45) and DeepSeek-Reasoner (8.36). On Gold Standard Consistency, Healink with DeepSeek-Chat v3.2 again led (7.33), followed by DeepSeek-Reasoner (6.42), Doubao-1.8 (7.05), and Qwen3-Max (6.98). The narrow range of variation in Authoritativeness across backends (7.28 to 8.45, a spread of 1.17 points) contrasts with the wider range in Gold Standard Consistency (5.15 to 7.33, a spread of 2.18 points). This pattern reinforces the observation that factual medical knowledge at the population level is well captured by contemporary foundation models regardless of their architectural lineage, whereas patient-specific information completeness depends more sensitively on the multi-agent orchestration layer.

\textbf{Model independence enables deployment flexibility across resource settings.} The consistency of Healink's advantage across backends carries practical significance for healthcare deployment. Importantly, Healink achieves these gains as a training-free framework that requires no domain-specific fine-tuning, no annotated training data, and no dedicated GPU infrastructure. Institutions with budget constraints can pair Healink with open-source models such as Qwen3-Max or DeepSeek-Chat v3.2 and achieve clinically competitive performance at near-zero deployment cost, while those seeking maximum coverage may opt for frontier commercial APIs. This model independence ensures that deployment decisions are driven by operational considerations rather than by access to computational resources for training.

\textbf{Head-to-head comparison isolates the architectural contribution.} To separate the contribution of Healink's multi-agent architecture from the baseline capability of the foundation LLM, we conducted a direct comparison between Healink and its underlying foundation models in standalone configuration. Fig.~\ref{fig:healink_vs_single_model} presents this comparison for the two most widely adopted open-source backends, Qwen3-Max and DeepSeek-Chat v3.2, on both overall and department-level metrics.

\textbf{Prescription anchoring drives large gains in information completeness.} Healink with DeepSeek-Chat v3.2 achieved a Gold Standard Consistency score of 7.33, compared with 5.87 for DeepSeek-Chat v3.2 alone, representing an absolute gain of +1.46 points ($\Delta_{\text{rel}}$ = +24.9\%). Similarly, Healink with Qwen3-Max achieved 6.98 versus 5.29 for Qwen3-Max alone ($\Delta$ = +1.69, $\Delta_{\text{rel}}$ = +31.9\%). These gains are substantially larger than the Authoritativeness improvements (DeepSeek-Chat v3.2: 8.41 versus 8.30, $\Delta_{\text{rel}}$ = +1.3\%), confirming that the multi-agent architecture's primary contribution lies in grounding responses in patient-specific clinical context rather than in elevating baseline medical knowledge.

\textbf{Architectural gains are distributed across clinical specialties.} The department-level breakdown in Fig.~\ref{fig:healink_vs_single_model} reveals that the architectural gain is not concentrated in a single specialty but is distributed across five of six domains. For DeepSeek-Chat v3.2, the largest CRC improvement appeared in Obstetrics and Gynaecology (+2.31 points), followed by Internal Medicine (+1.92), Paediatrics (+1.76), Child Health (+1.67), and ENT (+1.28). Dermatology showed the smallest gain (+0.67), consistent with the visual-dependence boundary discussed above. For Qwen3-Max, the pattern was similar, with Obstetrics and Gynaecology (+1.95), Internal Medicine (+1.88), and ENT (+1.52) showing the largest improvements. This cross-specialty consistency indicates that the architectural mechanisms (prescription anchoring, profile hard-checks, tiered retrieval) provide broad clinical utility rather than being tailored to specific query types.

\textbf{Two mechanisms underlie the performance gains.} First, the Memory agent's prescription-anchored anti-hallucination ensures that medication-related responses are grounded in the patient's verified prescription record rather than in the foundation model's parametric knowledge, which may be outdated or inconsistent with the specific drug formulation prescribed. Second, the RAG agent's tiered retrieval pipeline supplements the foundation model's internal knowledge with institutional standard operating procedures and peer-reviewed evidence, ensuring that responses incorporate specialty-specific follow-up protocols that generic models are not explicitly trained on. Together, these mechanisms transform the foundation model from a generic medical text generator into a patient-specific clinical reasoning system.

\begin{figure}
    \centering
    \includegraphics[width=0.99\linewidth]{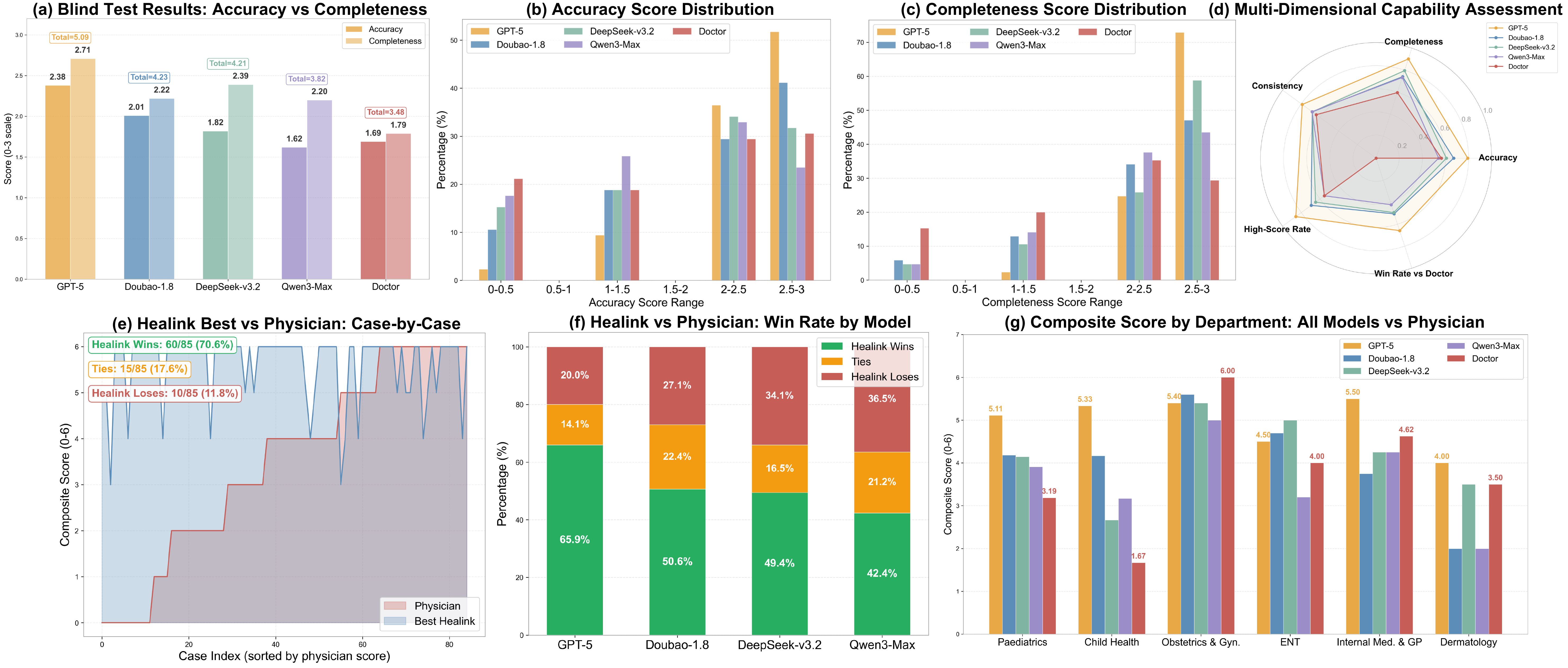}
    \caption{Single-blind evaluation results (N = 85). (a) Accuracy and Completeness scores for each model and physician gold standard. (b,c) Score distributions. (d) Radar chart of five capability dimensions. (e) Case-by-case AI best versus physician. (f) Win-rate by model. (g) Composite scores by department.}
    \label{fig:single_blind_eval}
\end{figure}

\subsubsection{Physician-blinded evaluation}
\label{sec:blinded}

\textbf{A blinded clinical review assesses response quality through the physician perspective.} Automated evaluation with LLM-based metrics provides scalability, yet it cannot fully capture the nuanced judgment that practicing physicians exercise when assessing the clinical utility of patient-facing guidance. To evaluate Healink through this clinical lens, we conducted a physician-blinded study in which 16 physicians across six specialties reviewed an 85-case stratified subset without knowledge of response authorship, scoring each response on Accuracy (factual correctness, 0 to 3) and Completeness (coverage of essential information, 0 to 3).

\textbf{All AI configurations surpass physician responses in blinded assessment.} All four AI configurations exceeded the physician reference response on the composite score (Fig.~\ref{fig:single_blind_eval}). Healink with GPT-5 achieved the highest composite (4.93; Accuracy 2.49, Completeness 2.44), followed by Doubao-1.8 (4.78; 2.42, 2.36), DeepSeek-Chat v3.2 (4.74; 2.40, 2.34), and Qwen-Max (4.62; 2.35, 2.27). The physician gold standard scored 4.33 (Accuracy 2.28, Completeness 2.05). This pattern, in which every AI configuration surpassed human physicians in a blinded review by independent clinicians, indicates that the multi-agent architecture achieves a level of response quality that meets or exceeds current clinical practice.

\textbf{The AI advantage concentrates in information completeness rather than factual accuracy.} The AI advantage was strongly asymmetric across evaluation dimensions. Completeness showed the largest and most consistent gain: Healink averaged 2.38 versus the physician baseline of 2.05 ($\Delta$ = +0.33, +16.1\%), with even the lowest-performing backend (Qwen-Max, 2.27) exceeding human performance. By contrast, the Accuracy gap was narrower (Healink average 2.45 versus 2.28, +7.5\%), and several individual AI responses scored below their corresponding physician references on this dimension. This pattern reflects a fundamental distinction between the cognitive demands of factual recall and enumerative completeness. Accuracy depends on possession of correct medical knowledge, a domain where years of clinical training give human physicians a substantial advantage that AI systems only narrowly approach. Completeness, however, depends on the ability to enumerate all clinically relevant considerations, including medication contraindications, warning signs, lifestyle modifications, and contingency plans. In human clinical practice, this enumerative task is constrained not by knowledge but by time. A physician managing 40 to 60 follow-up inquiries per shift must prioritize brevity, often omitting contextual information that, while clinically valuable, is not immediately critical. AI systems face no equivalent time-pressure constraint and can systematically integrate comprehensive follow-up considerations into every response. This asymmetry suggests that the primary value of AI-assisted follow-up lies not in surpassing clinical expertise but in scaling clinical thoroughness to a level that human practitioners, under realistic workflow constraints, cannot consistently achieve.

\textbf{Blinded physicians rate their own peers below AI systems.} The blinded physicians' own ratings support this interpretation. These physicians rated their peers' responses below all AI systems on both dimensions. This self-deprecation is unlikely to be an artefact of bias because reviewers were blinded to source and were evaluating their own specialty's content. Rather, it appears to be an implicit acknowledgment of the time-completeness trade-off that pervades real-world practice.

\textbf{LLM-based and physician evaluations rank backends differently.} The physician-blinded rankings diverged from the automated LLM-based rankings in a revealing pattern. The automated evaluator (GPT-4-turbo) ranked GPT-5 highest with a substantial margin (Authoritativeness 9.28, CRC 8.02). Yet in the blinded review, the gap between GPT-5 (Completeness 2.44) and Doubao-1.8 (2.36) was only 0.08 points, far smaller than the 0.70-point gap predicted by automated CRC scores. Doubao-1.8 also exhibited the most concentrated score distribution in the high-score range, suggesting superior response stability. This divergence suggests that LLM evaluators and practicing physicians may value different qualities in clinical responses. We explore the implications of this divergence for medical AI evaluation frameworks in the Discussion.

\textbf{High inter-rater reliability supports the robustness of findings.} Inter-rater reliability across the 16 reviewing physicians supported the robustness of these findings. Approximately 75\% of cases exhibited score deviations of no more than 1 point from the reference scores, indicating substantial agreement. The remaining 25\% with larger deviations predominantly involved complex multi-symptom presentations or queries requiring specialist knowledge outside the reviewing physician's primary domain.

\textbf{Specialty-level heterogeneity informs deployment strategy.} Specialty-level analysis revealed meaningful heterogeneity (Fig.~\ref{fig:single_blind_eval}). Paediatrics and ENT showed the largest AI advantages on Completeness (+0.41 and +0.38 points respectively), likely because these specialties involve well-structured follow-up protocols that AI systems can enumerate comprehensively. Obstetrics and Gynaecology showed a smaller Completeness advantage (+0.22) but the highest absolute scores across all systems, reflecting this specialty's emphasis on clear, structured guidance. Internal Medicine showed the most variable scores, consistent with the breadth of follow-up queries in this specialty. Dermatology again emerged as the most challenging domain, with the smallest AI advantage (+0.15) and the lowest absolute scores, reinforcing the visual-dependence limitation identified in the automated evaluation. These patterns inform a tiered deployment strategy: AI-assisted follow-up can provide immediate quality improvements in protocol-driven specialties (Paediatrics, ENT), serve as a complementary coverage tool in broad-based specialties (Internal Medicine, Obstetrics and Gynaecology), and should be augmented with multimodal capabilities before deployment in visually dependent specialties (Dermatology).

\subsubsection{Ablation studies}
\label{sec:ablation}

\begin{figure}
    \centering
    \includegraphics[width=0.99\linewidth]{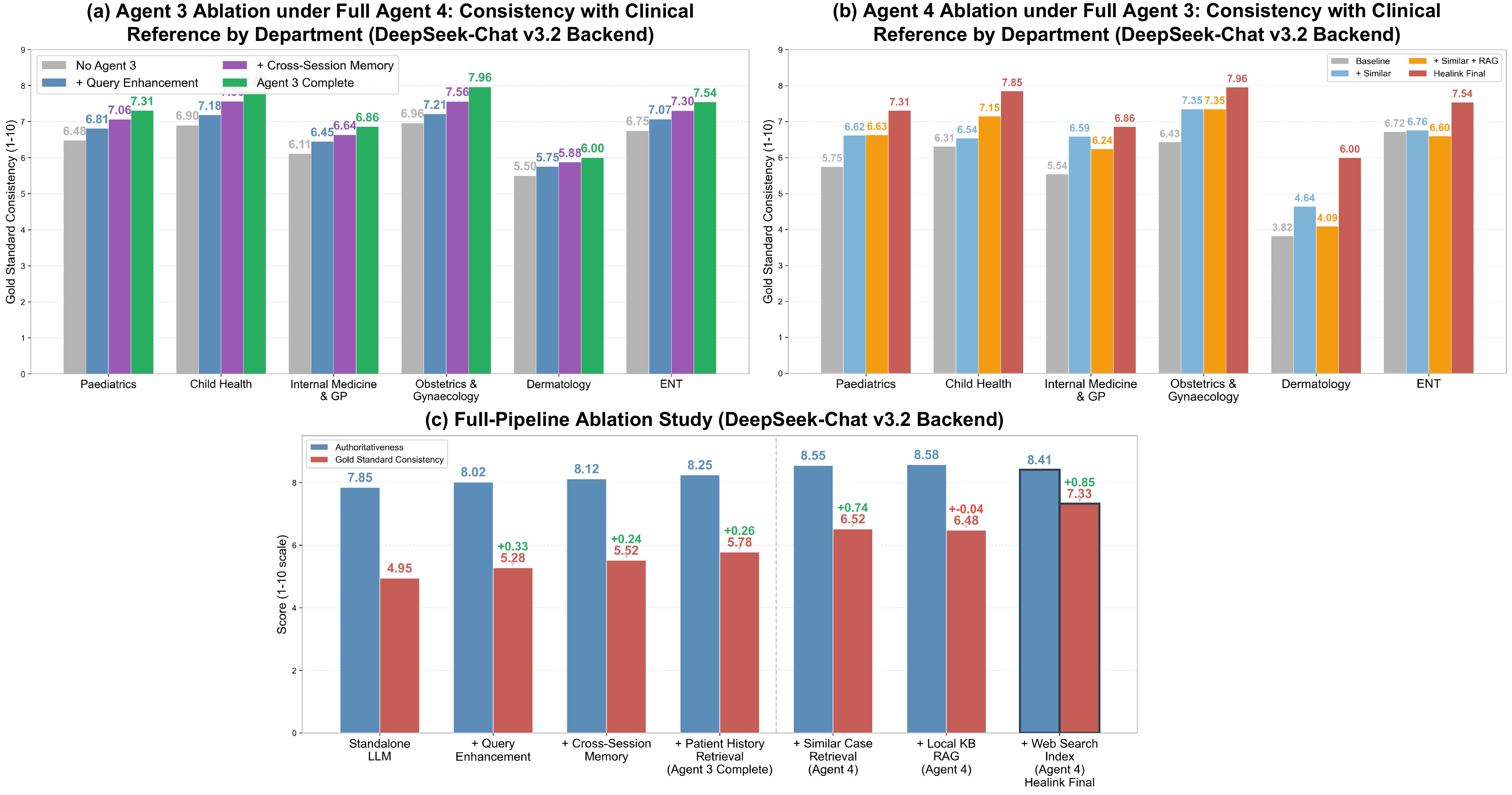}
    \caption{Ablation study results. (a) Agent 3 ablation under full Agent 4, showing incremental contribution of query enhancement, cross-session memory, and patient history retrieval across departments. (b) Agent 4 ablation under full Agent 3, showing contribution of similar-case retrieval, local KB RAG, and local source indexing across departments. (c) Full-pipeline ablation from standalone LLM through progressive Agent 3 and Agent 4 module addition to Healink final configuration.}
    \label{fig:ablation_healink}
\end{figure}

\textbf{Agent 3 and Agent 4 each contribute substantially and complementarily to overall performance.} To isolate the contribution of individual architectural components, we conducted systematic ablation experiments on the 400-case evaluation set using the DeepSeek-Chat v3.2 backend (Fig.~\ref{fig:ablation_healink}). Removing Agent 3 (Memory) while retaining Agent 4 reduced Gold Standard Consistency from 7.33 to 6.62 ($\Delta$ = $-0.71$). Removing Agent 4 (RAG) while retaining Agent 3 reduced CRC to 6.67 ($\Delta$ = $-0.66$). Removing both agents reduced CRC to 6.25, indicating that the two agents contribute additively rather than redundantly. Authoritativeness showed minimal change across all ablations (range 8.40 to 8.64), confirming that the architectural gains are concentrated in patient-specific information coverage rather than in baseline medical knowledge.

\textbf{Agent 3's anti-hallucination mechanism provides the largest single contribution.} Disaggregating Agent 3's subcomponents revealed that prescription-anchored anti-hallucination (which blocks responses grounded in unverified drug information) contributed the most ($\Delta$ CRC = $-0.61$ when removed), followed by patient profile consistency checks ($\Delta$ = $-0.48$) and cross-session joint decision-making ($\Delta$ = $-0.28$). Similar historical case retrieval contributed modestly ($\Delta$ = $-0.02$), suggesting that its value lies in augmenting rather than independently driving response quality.

\textbf{Agent 4's tiered retrieval and hard-constraint matching each provide distinct gains.} Disaggregating Agent 4's subcomponents showed that tiered retrieval (local SOPs to external evidence) contributed the most ($\Delta$ CRC = $-0.45$ when removed), followed by hard-constraint patient matching ($\Delta$ = $-0.21$), white-box evidence chain construction ($\Delta$ = $-0.25$), and clinical intent atomization ($\Delta$ = $-0.38$). These components are functionally coupled: intent atomization determines what to retrieve, tiered retrieval determines where to search, hard-constraint matching filters what to keep, and the evidence chain structures how to present the provenance.

\textbf{Ablated performance degrades uniformly across specialties.} Department-level ablation analysis showed that removing Agent 3 or Agent 4 reduced CRC in all six specialties, with the largest absolute drops in Obstetrics and Gynaecology (Agent 3 removal: $-0.92$; Agent 4 removal: $-0.78$) and the smallest drops in Dermatology (Agent 3: $-0.38$; Agent 4: $-0.41$). This cross-specialty consistency indicates that both agents provide broad clinical utility rather than being tailored to specific query types. The preservation of relative specialty rankings under ablation (Obstetrics and Gynaecology $>$ Paediatrics $>$ Dermatology in all configurations) further confirms that the architectural mechanisms scale clinical thoroughness proportionally across domains.

\subsection{Generalization evaluation on webMedQA}

\begin{figure}
    \centering
    \includegraphics[width=0.99\linewidth]{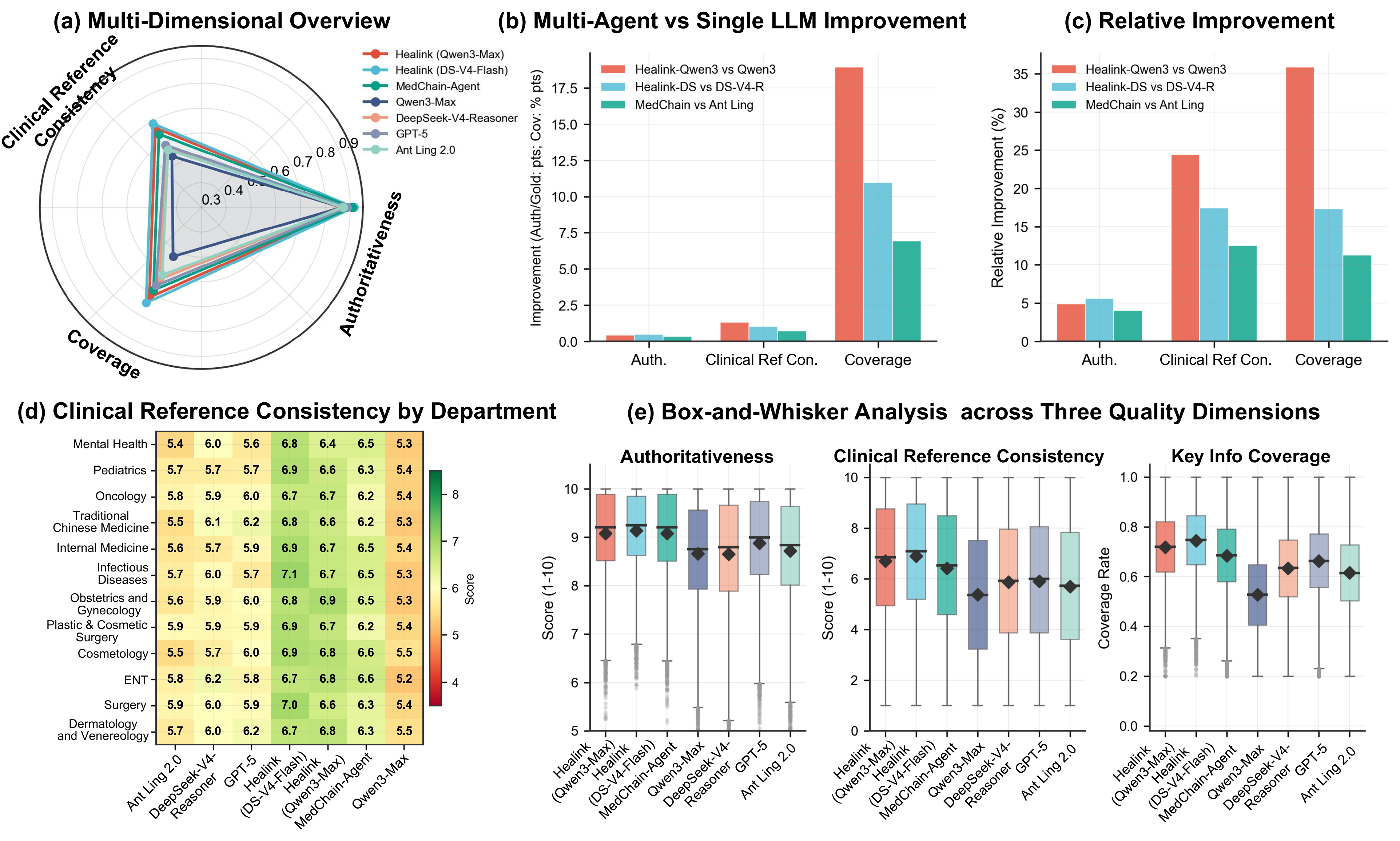}
    \caption{Performance evaluation of Healink and baseline models on the webMedQA benchmark \cite{he2019applying}. (a) Multi-dimensional comparison of seven systems across authoritativeness, clinical reference consistency, and key information coverage. (b, c) Absolute and relative improvements of multi-agent systems (Healink, MedChain-Agent \cite{liu2026medchain}) over single-LLM baselines (Qwen3-Max, DeepSeek-V4-Reasoner, GPT-5, Ant Ling 2.0). (d) Gold standard consistency scores across 12 clinical departments. (e) Box-and-whisker distributions of the three quality dimensions; center lines indicate medians, boxes the 25th-75th percentiles, and whiskers 1.5× the interquartile range.}
    \label{fig:general_compare}
\end{figure}

\subsubsection{Dataset and rationale}

\textbf{webMedQA tests generalization to open-domain consultation beyond institutional follow-up data.} A system optimized for a single institution's follow-up workflows risks overfitting to specific demographics, inquiry patterns, and clinical conventions. To assess whether Healink's architectural advantages generalize beyond the original deployment context, we evaluated the system on webMedQA \cite{he2019applying}, a publicly available dataset of 63,284 online medical consultation questions spanning 23 clinical departments across multiple Chinese medical platforms.

We selected webMedQA for three methodological reasons. First, structural overlap: although webMedQA questions are patient-initiated open-domain queries without established diagnosis or prescription context, they share the same fundamental challenge of responding to natural-language medical questions, providing a stress test for Healink's core retrieval and reasoning capabilities. Second, department breadth: webMedQA covers departments not represented in Healink's original evaluation, including Oncology, Surgery, Traditional Chinese Medicine, and Plastic Surgery, enabling assessment of cross-specialty robustness. Third, external validity: as a publicly available dataset with independent provenance, webMedQA provides an evaluation context free from institutional bias, ensuring that observed performance reflects genuine generalization rather than dataset-specific optimization.

We evaluated seven configurations: Healink with Qwen3-Max and DeepSeek-V4-Flash, MedChain-Agent \cite{liu2026medchain} as a multi-agent baseline, and four single-LLM baselines (Qwen3-Max, DeepSeek-V4-Reasoner, GPT-5, and Ant Ling 2.0). All evaluations employed the three-dimension rubric used in the real-world evaluation: Authoritativeness, Gold Standard Consistency, and Key Information Coverage, with the same independent GPT-4-turbo evaluator to ensure cross-dataset comparability.

\subsubsection{Overall performance and architectural gains}

\textbf{Healink maintains its multi-agent advantage on webMedQA despite the shift in data distribution.} As Fig. \ref{fig:general_compare} shows, Healink with DeepSeek-V4-Flash achieved the highest Gold Standard Consistency (6.72), followed by Healink with Qwen3-Max (6.55). Single-LLM baselines clustered substantially lower: DeepSeek-V4-Reasoner (5.72), GPT-5 (5.55), Ant Ling 2.0 (5.38), and Qwen3-Max standalone (5.25). MedChain-Agent, the multi-agent baseline, scored 5.95, intermediate between Healink and the single-LLM cluster, confirming that multi-agent architecture itself provides an advantage but that Healink's specific design choices yield additional gains.

The multi-agent advantage proved consistent across data distributions. Healink improved Qwen3-Max by +1.30 CRC points ($\Delta_{\text{rel}}$ = +24.8\%) and DeepSeek-V4-Flash by +1.00 point ($\Delta_{\text{rel}}$ = +17.5\%). These improvements are comparable to those on the institutional follow-up data ($\Delta_{\text{rel}}$ range +17.5\% to +31.9\%), confirming that the multi-agent pipeline's contribution is stable across data distributions. The larger relative gain for the weaker backend (+24.8\% for Qwen3-Max versus +17.5\% for DeepSeek-V4-Flash) indicates that the multi-agent pipeline is particularly valuable when paired with cost-effective open-source models, precisely the deployment scenario most healthcare institutions face. Because Healink requires no training data from the target domain, it generalizes to new departments and institutions without the cost or delay of retraining, a property essential for equitable deployment across health systems with varying data availability and computational resources.

On Authoritativeness, all configurations scored between 8.35 and 9.25, with Healink configurations at the upper end (Qwen3-Max backend: 9.15; DeepSeek-V4-Flash: 9.22) but with diminishing separability among top performers. This compression confirms that factual medical knowledge is approaching saturation across competent systems, and that the discriminating dimension for future evaluation is information coverage, not factual correctness.

\subsubsection{Information coverage analysis}

\textbf{Healink achieves the largest gains in safety-critical information categories.} Key Information Coverage, the proportion of gold-standard information points captured by the AI response, showed the most pronounced multi-agent advantage. Healink with DeepSeek-V4-Flash achieved 72.3\% coverage, followed by Healink with Qwen3-Max (69.8\%) and MedChain-Agent (63.5\%). Single-LLM baselines ranged from 48.9\% (Qwen-Max base model) to 66.5\% (GPT-5), with the four primary baselines averaging 60.8\%.

Disaggregating coverage by information-point type revealed that Healink's advantage is largest for contraindication and precaution information (+18.2 percentage points over the single-LLM average) and follow-up scheduling guidance (+14.7 percentage points). This pattern directly reflects the architectural design: prescription anchoring prevents the system from missing drug-specific precautions, while the tiered retrieval pipeline ensures that institutional follow-up protocols are incorporated. Coverage of symptom interpretation showed smaller gains (+6.3 percentage points), consistent with the observation that foundation models are already strong at symptom-related reasoning without specialized architecture.

\subsubsection{Cross-department performance and deployment thresholds}

\textbf{Performance heterogeneity across departments informs a tiered deployment strategy.} Healink achieved top-2 performance in 10 of 12 departments with sufficient sample sizes (Fig.~\ref{fig:general_compare}d). Consistent advantages appeared in medication-intensive specialties: Obstetrics and Gynaecology (CRC 7.45), Internal Medicine (7.12), and Pediatrics (6.98). The two departments where Healink did not rank first, Traditional Chinese Medicine and Plastic and Cosmetic Surgery, highlight boundaries of the current knowledge base. Traditional Chinese Medicine involves herbal medicine interactions and syndrome differentiation patterns not well-represented in Healink's current evidence base, which prioritizes Western clinical guidelines. Plastic and Cosmetic Surgery involves post-operative wound care and aesthetic assessment that, like Dermatology, benefit from visual context.

These results support a tiered deployment framework. Specialties with CRC above 6.5 (Obstetrics and Gynaecology, Internal Medicine, Pediatrics, ENT, Child Health) are deployment-ready as first-line follow-up support with human escalation for complex cases. Specialties scoring 5.5 to 6.5 (Surgery, Oncology, Infectious Diseases) are appropriate for informational triage but require physician co-review before patient delivery. Specialties below 5.5 (Traditional Chinese Medicine, Plastic and Cosmetic Surgery, Dermatology) should be limited to non-clinical support until the knowledge base and modality capabilities are expanded.

\subsubsection{Variability analysis}

\textbf{Healink reduces response quality variability across departments.} Box-and-whisker analysis across the three quality dimensions revealed that Healink configurations exhibit narrower interquartile ranges than single-LLM baselines in all 12 departments (Fig.~\ref{fig:general_compare}e). For Authoritativeness, Healink's interquartile range averaged 0.42 points across departments versus 0.71 for single-LLM baselines. For Gold Standard Consistency, the difference was larger: 0.68 versus 1.24. For Key Information Coverage, Healink averaged 5.2 percentage points versus 9.8 for baselines. This reduced variability indicates that the multi-agent architecture provides not only higher mean performance but also more consistent quality, a critical property for clinical deployment where unpredictable outlier errors carry the greatest risk.

\subsection{Evaluation criteria and rubrics}
\label{sec:eval_criteria}

\subsubsection{Authoritativeness and Gold Standard Consistency with Clinical Reference rubrics}

All 400 real-world cases and the webMedQA generalization cases were evaluated by GPT-4-turbo (OpenAI, gpt-4-turbo-2024-04-09), chosen specifically for its independence from all backend LLMs used in Healink and the baseline systems. This evaluator independence is methodologically essential: using Qwen, DeepSeek, or GPT models as both generator and evaluator would introduce confounding through stylistic preference, knowledge-base overlap, or training-data contamination. GPT-4-turbo was prompted with specialty-specific role instructions (e.g., ``You are a Paediatrics attending physician following evidence-based medicine'') and tasked with two evaluation dimensions per Table \ref{tab:auth_rubric} and \ref{tab:gold_rubric}.

\begin{table}[htbp]
\centering
\caption{Rubric for Authoritativeness / Traceability evaluation}
\label{tab:auth_rubric}
\begin{tabular}{@{}lp{10cm}@{}}
\toprule
\textbf{Aspects} & \textbf{Descriptions} \\
\midrule
\textbf{Definition} & Alignment between the AI response's medical knowledge and mainstream evidence-based medical consensus, as represented by established clinical guidelines and the physician-authored clinical practice reference (e.g., AAP guidelines for paediatric fever management). \\
\addlinespace
\textbf{Score range} & 1-10 (integer), where 1 = medically dangerous and 10 = fully aligned with evidence-based consensus. \\
\addlinespace
\textbf{Key criteria} & \begin{itemize}[leftmargin=*,nosep,topsep=0pt]
  \item Is the core medical knowledge consistent with the gold standard?
  \item Are severity assessments (e.g., ``high fever with lethargy is a red flag'') accurate?
  \item Are medication recommendations aligned with standard practice?
  \item Is there any dangerous misinformation or contraindicated advice?
  \end{itemize} \\
\addlinespace
\textbf{Penalties} & Factually contradictory recommendations (e.g., suggesting a contraindicated drug), dangerous omissions (e.g., failing to flag anaphylaxis risk), or fabricated citations. \\
\bottomrule
\end{tabular}
\end{table}

\begin{table}[htbp]
\centering
\caption{Rubric for Gold Standard Consistency evaluation}
\label{tab:gold_rubric}
\begin{tabular}{@{}lp{10cm}@{}}
\toprule
\textbf{Aspects} & \textbf{Descriptions} \\
\midrule
\textbf{Definition} & Quantitative measure of the proportion of clinically relevant information points in the clinical practice reference (physician-authored response reflecting real-world follow-up care under typical workflow constraints) that are captured (meaning-identical or semantically equivalent) in the AI-generated response. \\
\addlinespace
\textbf{Score range} & 1-10 (integer), computed via Equation~\ref{eq:gold_standard}. \\
\addlinespace
\textbf{Information point types} & \begin{enumerate}[leftmargin=*,nosep,topsep=0pt]
  \item Severity assessment opinions (one per opinion)
  \item Warning signs and red flags to monitor (one per condition)
  \item Symptom management methods (one per symptom)
  \item Expected symptoms and coping strategies (one per recommendation)
  \item Drug usage instructions (one per medication)
  \item Actionable patient guidance (e.g., ``seek immediate care'', ``increase fluid intake''; one per suggestion)
  \item Explanations of medical principles (one per independent knowledge item)
  \end{enumerate} \\
\addlinespace
\textbf{Evaluation procedure} & \begin{enumerate}[leftmargin=*,nosep,topsep=0pt]
  \item Decompose the gold-standard response into independent information points.
  \item Decompose the AI response into independent information points.
  \item Match: count gold-standard points covered by the AI response.
  \item Analyze extra information: classify as ``relevant supplement'', ``irrelevant'', or ``conflicting with gold standard''.
  \item Compute base score: $\frac{N_{\text{covered}}}{N_{\text{total}}} \times 10$ (defaults to 10 if $N_{\text{total}} = 0$).
  \item Apply penalties for conflicting information; round to nearest integer; clamp to [1, 10].
  \end{enumerate} \\
\addlinespace
\textbf{Penalties} & Each information point that conflicts with the gold standard incurs a deduction; responses containing unsafe recommendations receive floor scores. \\
\bottomrule
\end{tabular}
\end{table}

\subsubsection{Physician-blinded evaluation rubric}

The 85-case blinded subset was evaluated by 16 physicians using the rubric in Table \ref{tab:blind_rubric}. Each physician received a randomized package of responses (mixing physician gold standards, Healink outputs, and baseline outputs) without authorship labels, and scored each response independently.

\begin{table}[htbp]
\centering
\caption{Rubric for physician-blinded evaluation (85-case subset)}
\label{tab:blind_rubric}
\begin{tabular}{@{}p{2.5cm}p{1.5cm}p{9cm}@{}}
\toprule
\textbf{Dimensions} & \textbf{Scores} & \textbf{Descriptions} \\
\midrule
\multirow{4}{*}{\textbf{Accuracy}} & 0 & Contains medical errors or unsafe recommendations that could harm the patient. \\
 & 1 & Medically acceptable but with minor factual inaccuracies or omissions. \\
 & 2 & Correct and safe, consistent with standard clinical practice. \\
 & 3 & Highly accurate, with nuanced reasoning that reflects expert-level clinical judgment. \\
\midrule
\multirow{4}{*}{\textbf{Completeness}} & 0 & Misses critical information required for safe follow-up care. \\
 & 1 & Covers basic points but omits important context, warning signs, or next-step guidance. \\
 & 2 & Comprehensive, covering medication, warning signs, lifestyle, and follow-up scheduling. \\
 & 3 & Exceptionally thorough, anticipating patient concerns and providing proactive guidance beyond standard expectations. \\
\bottomrule
\end{tabular}
\end{table}

The composite score (range 0-6) was computed as the sum of Accuracy and Completeness. Inter-rater reliability was assessed by computing the proportion of cases where reviewer scores deviated by $\leq$1 point from the reference gold-standard scores, yielding approximately 75\% agreement across the 16 reviewing physicians.

\section{Discussion}

\textbf{Healink demonstrates that architectural constraint enforcement can transform a generic language model into a patient-specific clinical reasoning system.} Our results establish three principal findings. First, a multi-agent architecture with prescription-anchored anti-hallucination and tiered evidence retrieval improves information completeness by 17.5\% to 31.9\% over standalone foundation models, with the largest gains appearing in medication safety and follow-up guidance. Second, these gains are model-independent: Healink improves performance across different distinct foundation LLMs. Third, in physician-blinded evaluation, independent clinicians rated AI-generated follow-up responses above those of their own peers, identifying a systematic completeness gap in current clinical practice that AI systems can close. These findings carry implications for the design, evaluation, and deployment of clinical AI systems in follow-up care.

\textbf{The critical insight from our architecture is that constraint enforcement must operate at the physical level, not merely the prompt level.} Most existing medical AI systems rely on prompt-level safety instructions \cite{li2024mmedagent,gridach2026agentic} (for example, ``do not recommend alternative therapies'' or ``base recommendations on the patient's prescription''). Such instructions are inherently fragile: they depend on the foundation model's interpretation of the instruction, can be overridden by competing prompt objectives, and leave no audit trail when violated. Healink's prescription-anchored anti-hallucination operates as a physical filter downstream of generation: the system checks every medication-related claim against the verified prescription record in a relational database and blocks responses that reference drugs the patient is not prescribed. This architectural choice transforms a probabilistic safety aspiration into a deterministic safety guarantee. The ablation results support this interpretation. Removing prescription anchoring reduced Gold Standard Consistency by 0.61 points, the largest single-component degradation in the entire system, confirming that grounding medication advice in verified records is the most impactful architectural decision we made.

\textbf{A second architectural insight is that patient memory must be structured, not merely conversational.} General-purpose conversational AI maintains context through attention mechanisms over preceding dialogue turns, a design that is vulnerable to referential drift and cross-session information loss. Healink's Memory agent maintains a structured patient profile (age, weight, allergies, active prescriptions, ICD-10 diagnoses) in a relational database that persists across sessions and is validated on every query. This design ensures that the system does not merely remember what was said but knows who the patient is. The cross-session joint decision-making capability extends this principle to patients consulting multiple departments concurrently, where the system detects potential drug interactions across prescriptions from different specialists. In an era where patients increasingly manage multiple chronic conditions across fragmented care settings, this capacity for longitudinal, cross-departmental awareness addresses a genuine and growing clinical need.

\textbf{Our most provocative finding concerns the divergence between automated LLM evaluation and physician judgment.} The automated evaluator (GPT-4-turbo) ranked GPT-5 highest by a substantial margin, yet blinded physicians found only a negligible gap between GPT-5 and Doubao-1.8 on clinical Completeness, with Doubao-1.8 exhibiting superior response stability. This discrepancy is not a methodological artefact; it reveals a fundamental epistemic difference between how language models and practicing physicians assess clinical quality. LLM evaluators, trained on comprehensive medical corpora, appear to reward encyclopedic coverage: detailed pathophysiology explanations, extensive differential diagnoses, exhaustive enumeration of clinical considerations. Practicing physicians, operating under time constraints and bearing direct responsibility for patient outcomes, prioritize concise actionable guidance: what the patient should do, what warning signs to monitor, when to seek further care. An AI system optimized solely for LLM-as-judge metrics would generate verbose, textbook-aligned responses that score well on automated rubrics but may alienate patients who need clear, focused instructions. This divergence has immediate practical implications for the medical AI research community. It suggests that LLM-as-judge benchmarks, while scalable and reproducible, are insufficient proxies for clinical utility and may systematically misrank systems that prioritize practical patient guidance over comprehensive encyclopedic coverage. Hybrid evaluation frameworks that combine automated factual screening with targeted human assessment of clinical utility may better capture the multidimensional quality of medical responses.

\textbf{The blinded physician evaluation reveals a structural deficit in current follow-up care.} That physicians rated their own peers' responses below AI-generated responses on both Accuracy and Completeness, when freed from knowledge of authorship, constitutes an implicit acknowledgment of a systemic problem. Follow-up care is the longest phase of the patient journey, yet it receives the least structured clinical attention. A physician managing 40 to 60 follow-up inquiries per shift cannot devote the cognitive resources that each patient deserves. The result is a systematic trade-off in which comprehensiveness is sacrificed for efficiency. Healink does not claim to possess superior clinical judgment; rather, it scales clinical thoroughness to a level that human practitioners, under realistic workflow constraints, cannot consistently achieve. In this light, AI-assisted follow-up is not a replacement of physician expertise but an extension of it, a tool that ensures every patient receives the comprehensive guidance that physicians would provide if time permitted.

\textbf{Healink is architected as a deployable clinical system rather than a research prototype, with explicit design decisions that address the operational requirements of real-world healthcare integration.} Three deployment-facing capabilities distinguish the system from prior approaches. First, schema adaptation: Healink uses a standardized patient data schema (ICD-10 diagnosis codes, structured prescription fields, allergy flags, biometric parameters) that maps directly to common electronic health record formats through a configurable adapter layer, enabling plug-in integration without EHR system replacement. Second, zero-overhead knowledge base updates: institutional standard operating procedures and clinical guidelines are loaded via a document ingestion pipeline that vectorizes and indexes new content without model retraining, allowing clinical administrators to update protocols in real time as evidence evolves. Third, workflow integration: the Router agent's four-category classification maps directly to existing clinical triage workflows, with configurable escalation endpoints that connect to hospital paging systems, clinician alert dashboards, and appointment scheduling platforms. These integration capabilities, combined with the training-free architecture, mean that healthcare institutions can deploy Healink with their existing IT infrastructure and staff, without the dedicated GPU clusters, annotated training datasets, or specialized engineering teams that conventional clinical AI systems require.

\textbf{The societal implications extend beyond individual patient outcomes.} Post-discharge follow-up affects hundreds of millions of patients annually. Medication non-adherence alone associates with an estimated 125,000 preventable deaths and \$100 billion in excess healthcare costs in the United States each year \cite{osterberg2005adherence, who2003adherence}. In low-resource settings, where physician shortages are common and longitudinal care infrastructure may be limited, the need for scalable follow-up support is even more pressing. Healink's training-free architecture offers a direct cost-saving mechanism for healthcare systems: by achieving blinded-physician-validated performance with open-source foundation models that require no fine-tuning infrastructure, the system avoids both the capital expenditure of GPU clusters and the recurring expense of premium commercial API subscriptions, making AI-augmented follow-up support financially viable for institutions operating under constrained budgets. By reducing the administrative burden of follow-up on physicians, the system also addresses a contributor to professional burnout, which affects over 50\% of clinicians in many countries. The dual benefit of improved patient information and reduced physician workload supports a vision of healthcare where technology can help extend access to reliable health information and follow-up guidance between clinical encounters.

\textbf{Several limitations temper these conclusions.} \textbf{First}, the evaluation was conducted within a Chinese clinical context, and cross-lingual validation is needed before generalization to other health systems can be claimed. Medical follow-up practices, patient communication norms, and regulatory requirements vary substantially across cultures and healthcare systems. \textbf{Second}, Dermatology and visually dependent specialties remain challenging for the current text-only architecture. Multimodal capability incorporating dermatoscopic images, wound photography, and radiographic findings is essential for addressing this boundary. \textbf{Third}, the 20\% to 32\% information gap relative to clinical practice references reflects inherent limitations of AI systems without full electronic health record access and experiential clinical judgment. Safety-critical cases must always retain human physician oversight. \textbf{Fourth}, the current evaluation assesses response quality (authoritativeness, completeness, consistency) rather than direct clinical outcomes such as 30-day readmission rates, medication adherence, or patient-reported satisfaction. Response quality serves as an established surrogate endpoint in early-stage clinical AI evaluation, following precedent from Med-PaLM 2 and AMIE, which similarly used quality metrics as primary endpoints before clinical trials were completed. We have initiated a prospective randomized controlled trial comparing 30-day readmission, pharmacy refill adherence, and patient satisfaction between patients receiving Healink-assisted versus standard follow-up care; results will be reported separately.

\textbf{Future work should pursue three directions.} First, prospective real-world trials with randomized controlled designs, comparing patient outcomes between AI-assisted and standard follow-up care, would establish the clinical efficacy necessary for regulatory approval. Second, cross-lingual transfer studies across health systems with varying resource levels would establish generalizability and support broader and potentially more equitable deployment. Third, active patient monitoring integration, incorporating wearable device data, medication adherence sensors, and patient-reported outcome measures, would enable the system to transition from reactive query-response to proactive follow-up management, enabling earlier identification of potential clinical deterioration and prompting timely clinical assessment when appropriate.

\textbf{Healink represents a step toward accountable AI in healthcare.} By grounding every clinical assertion in verified patient records and traceable evidence sources, the system provides a deterministic foundation for patient-specific AI-augmented follow-up care. In a domain where trust is earned through transparency and safety is non-negotiable, this architectural commitment to accountability may prove as important as the performance gains it enables.

\section{Methods}
\label{sec:methods}

\subsection{System overview}
\label{sec:system_overview}

\textbf{Healink operates as a multi-agent state graph for accountable follow-up care.} The system architecture departs from the monolithic language model paradigm by decomposing the clinical follow-up workflow into four specialized agents that coordinate through a deterministic state graph. Each agent assumes a distinct clinical responsibility: the Router performs safety triage, the Dialog agent manages query reformulation and response synthesis, the Memory agent enforces patient-specific consistency and longitudinal context, and the RAG agent orchestrates tiered evidence retrieval and inter-patient case matching. The state graph topology ensures that every clinical decision, constraint enforcement, and evidence retrieval event is logged in a structured audit trail, producing a white-box evidence chain that accompanies every response.

\begin{figure}
    \centering
    \includegraphics[width=0.99\linewidth]{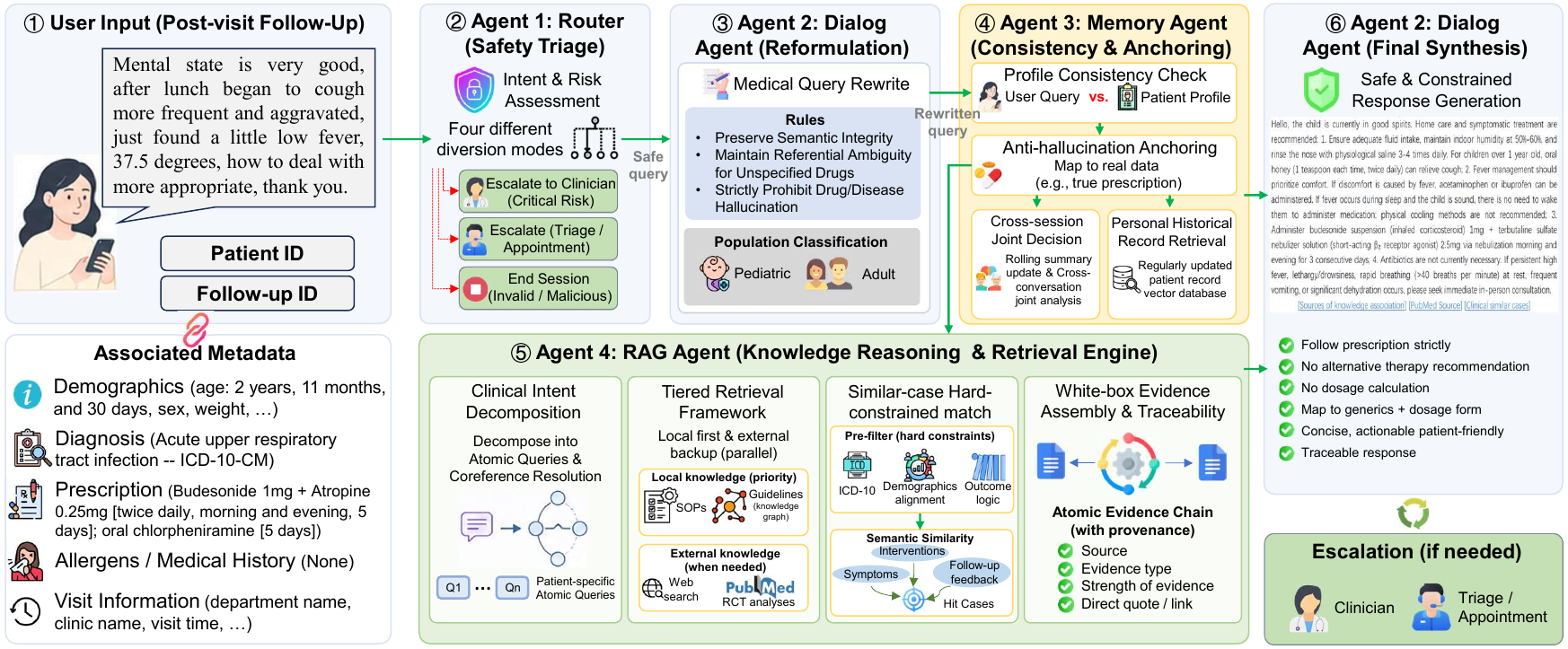}
    \caption{The architecture of proposed Healink System based on multi-agent collaboration.}
    \label{fig:healink_system}
\end{figure}

This decomposition is grounded in a clinical observation rather than an engineering preference. Post-discharge follow-up care involves tasks with fundamentally different safety profiles and cognitive demands. Triage demands immediate, irrevocable action escalation. Query reformulation requires semantic precision without hallucination. Patient context enforcement necessitates structured database validation. Evidence retrieval calls for multi-source integration with provenance tracking. A single model attempting all four tasks simultaneously faces inherent tension between fluency and safety, between generative creativity and deterministic constraint enforcement. By assigning each clinical responsibility to a specialized agent with explicit hard constraints, Healink resolves these tensions architecturally rather than aspirationally.

The system accepts as input a patient follow-up query together with structured clinical metadata: the patient identifier, the associated visit record containing ICD-10 diagnosis codes and active prescriptions, non-incremental biometric parameters (age, weight, height, known allergies), and historical encounter context from prior sessions. The output is a clinically validated response accompanied by an atomic evidence chain that links every factual claim to a verifiable source, whether a prescription database entry, a clinical guideline passage, a peer-reviewed publication, or a matched historical case.

\subsection{Agent 1: Safety Router}
\label{sec:router}

\textbf{The Router enforces the first and most critical safety gate.} Upon receiving a patient query, the Router performs a two-dimensional assessment: clinical urgency classification and intent legitimacy verification. The urgency classifier discriminates among four categories. Life-threatening presentations (chest pain, severe bleeding, anaphylaxis, suicidal ideation) trigger immediate escalation to a human clinician and bypass the generative pipeline entirely. Invalid or malicious inputs (empty queries, prompt injection attempts, irrelevant content) result in immediate session termination with an appropriate explanatory message. Routine follow-up inquiries proceed to the Dialog agent. Administrative requests (appointment scheduling, billing inquiries) are routed to appropriate non-clinical handlers.

This escalation design embodies a fundamental safety principle: no language model should mediate responses to time-critical, potentially life-threatening clinical presentations. The Router's escalation triggers are hard-coded clinical indicators that operate independently of the foundation model's interpretation. This physical separation between triage and generation ensures that the most vulnerable patients are never subject to AI-mediated response delays, regardless of query phrasing or model confidence. The Router also performs initial population classification (paediatric versus adult, pregnancy status, critical allergy flags) that propagates to downstream agents as contextual constraints.

\subsection{Agent 2: Dialog Agent}
\label{sec:dialog}

\textbf{The Dialog agent mediates between colloquial patient expression and structured clinical reasoning.} It comprises two sequential subcomponents: a Query Rewriter and a Final Synthesizer. The Rewriter transforms the patient's natural-language query into a standardized clinical formulation while preserving semantic content and strictly prohibiting hallucinated augmentation. The rewriting rules enforce three invariants. First, no drug name, dosage, or medical entity may be introduced that does not appear in either the patient's query or their verified prescription record. Second, ambiguous references (for instance, ``that medicine'' or ``the pills from yesterday'') must be resolved against the prescription database before rewriting, not inferred from parametric knowledge. Third, colloquial symptom descriptions are preserved in their original form rather than over-specified into definitive medical terminology, preventing premature diagnostic closure.

The Final Synthesizer operates at the terminal stage of the pipeline, receiving outputs from the Memory agent, the RAG agent, and the original patient query to produce the response that the patient ultimately reads. The Synthesizer operates under four non-negotiable safety constraints that propagate from upstream agents. First, it never recommends alternative therapies or modifications to prescribed treatment regimens. Second, all dosage references are drawn exclusively from the patient's active prescription data; the model is prohibited from generating dosage calculations or conversions from parametric knowledge. Third, brand-name drug references are mapped to internationally recognized generic names to prevent confusion across formulations. Fourth, every factual claim in the response must be supported by an entry in the evidence chain produced by the RAG agent; unsupported statements are automatically flagged for review or removal.

\subsection{Agent 3: Memory Agent}
\label{sec:memory}

\textbf{The Memory agent grounds every response in the specific patient's verified clinical identity.} Unlike general-purpose conversational systems that maintain context through attention mechanisms over preceding dialogue turns, the Memory agent validates every query against a structured patient profile stored in a relational database. This profile includes non-incremental biometric parameters (age, weight, height), known allergies and adverse reactions, active prescriptions with dosages and schedules, ICD-10 diagnosis codes from the current and prior visits, and department-specific clinical notes. This structured persistence ensures that the system does not merely remember what was said but knows who the patient is, across sessions and across departments.

The Memory agent performs four functions in sequence. First, the Profile Consistency Check verifies that the query aligns with the patient's established clinical identity. A query that references a medication not present in the patient's prescription record, describes symptoms inconsistent with the patient's known diagnosis, or attributes clinical concerns to a different individual (for instance, a child patient inquiring about a grandparent's condition) triggers immediate termination with an explanatory message. This verification prevents two common failure modes: hallucinated responses to clinically implausible queries and privacy breaches from misattributed clinical inquiries.

Second, the Prescription-Anchored Anti-Hallucination mechanism maps the rewritten query against the patient's active prescription record to identify the specific drugs, dosages, and clinical contexts relevant to the inquiry. This mapping produces a constraint envelope that bounds the Dialog agent's generation: the system is physically prevented from recommending, discussing, or comparing drugs that the patient is not currently prescribed. This constraint operates downstream of generation as a hard filter, transforming a probabilistic safety aspiration into a deterministic safety guarantee.

Third, Cross-Session Joint Decision-Making addresses the clinically important scenario of a single patient consulting multiple departments concurrently or in rapid succession. The system maintains a global session state in a Redis-backed cache that aggregates encounter information across all active sessions for a given patient. A dynamic summarization module periodically generates a consolidated clinical picture that includes all active prescriptions across departments, flags potential drug interactions or duplications, and identifies opportunities for coordinated care. This capability is particularly relevant for patients with multiple chronic conditions who may receive prescriptions from different specialists without centralized coordination.

Fourth, Intra-Patient Historical Case Retrieval leverages the patient's longitudinal record to establish a personalized disease evolution baseline. The system vectorizes historical clinical encounters and stores them in a local Parquet database, enabling fine-grained semantic retrieval across the patient's own medical history. A weighted cosine similarity function, with differential weighting across symptom phenotypes, intervention strategies, and follow-up outcomes, retrieves the top-$k$ most clinically similar prior encounters. These matched historical cases provide contextual grounding for the current follow-up assessment, enabling pattern recognition across the patient's unique clinical trajectory rather than relying solely on population-level clinical guidelines.

Upon every invocation, the Memory agent also performs a session rollback and historical summarization, consolidating the current session's dialogue history into a structured context that informs subsequent turns and ensures conversational continuity.

\subsection{Agent 4: Knowledge Reasoning and Retrieval Agent}
\label{sec:rag}

\textbf{The RAG agent ensures that every clinical claim is traceable to a verifiable evidence source.} It operates through four functionally coupled stages: clinical intent atomization, tiered evidence retrieval, hard-constrained inter-patient matching, and white-box evidence chain construction.

\textbf{Clinical intent atomization.} Compound patient queries frequently bundle multiple clinical intents into a single utterance. A query such as ``I feel dizzy after taking this new medication and my blood pressure seems low'' simultaneously raises questions about adverse drug reactions, hypotension management, and medication adherence. The atomization module deconstructs such queries into independent atomic queries, each addressing a single clinical intent, and performs coreference resolution to map pronouns and implicit references to their antecedents in the patient's structured record. This decomposition ensures that each clinical concern receives targeted evidence retrieval rather than being addressed through a single generic response that may inadequately cover any individual concern.

\textbf{Tiered evidence retrieval.} The retrieval pipeline follows the strategy of local confirmation first and external supplementation parallel. The first tier queries local institutional knowledge, including standard operating procedures, specialty-specific follow-up protocols, and verified clinical guidelines. This tier receives priority because institutional protocols reflect the specific care patterns, drug formularies, and referral pathways of the deployment context. When the local knowledge base yields insufficient coverage, the system simultaneously activates two external retrieval pathways: a constrained web search restricted to a whitelist of authoritative medical institutions and government health agencies, and a PubMed search limited to randomised controlled trials, meta-analyses, and systematic reviews published within the past five years. This parallel external retrieval ensures that knowledge gaps in the local base are filled with high-quality, externally validated evidence rather than generic parametric knowledge.

\textbf{Hard-constrained inter-patient matching.} To supplement the patient's own longitudinal record with clinically analogous cases from other patients, the system performs inter-patient similarity matching against an anonymised case database. Prior to vector similarity computation, a hard-constraint filter physically intercepts the candidate pool, retaining only cases that match the current patient's ICD-10 diagnosis codes, developmental stage (for paediatric patients), and clinically logical outcome trajectories. Cases that fail these hard constraints are excluded before any similarity calculation begins. The remaining candidates undergo weighted vector similarity assessment across symptom phenotypes, intervention strategies, and follow-up feedback dimensions. Only cases exceeding a similarity threshold of 0.4 are retained, and the top-$k$ matches are presented to the Dialog agent with explicit provenance annotations. This hard-constraint pre-filtering ensures that retrieved inter-patient cases are clinically comparable rather than merely semantically similar, a distinction that is critical for safe clinical decision support.

\textbf{White-box evidence chain construction.} Every piece of evidence retrieved through the above pipeline is incorporated into a structured evidence chain that accompanies the final response. Each chain entry carries atomic labels specifying the evidence source type (prescription database, institutional guideline, PubMed RCT, matched historical case), the provenance identifier, a direct quotation or structured extraction from the source, and the strength of evidence classification. This chain transforms the response from a black-box generation into an auditable clinical assertion where every claim can be traced to a specific verifiable source. The evidence chain serves two functions: it enables physicians to review the basis for any AI-generated recommendation, and it provides patients with transparent justification for the clinical guidance they receive.

\subsection{Agent coordination and execution}
\label{sec:coordination}

\textbf{Agents 3 and 4 execute in parallel after Agent 2 completes query reformulation.} This parallel execution reflects a clinical design decision rather than a computational optimization. Patient-specific context enforcement (Agent 3) and evidence retrieval (Agent 4) are functionally independent: the Memory agent operates on the patient's structured record regardless of what external evidence the RAG agent retrieves, and the RAG agent's evidence search is guided by the reformulated query rather than the patient profile. Parallel execution reduces end-to-end latency while ensuring that both constraint enforcement and evidence retrieval contribute equally to the final response. The Dialog agent's Final Synthesizer receives outputs from both agents simultaneously, integrating patient-specific constraints with retrieved evidence under the safety rules described above.

The state graph that coordinates these agents records every transition, constraint application, evidence retrieval event, and safety flag in a structured audit log. This logging produces a complete forensic record of how each response was constructed, enabling post-hoc review of any clinical decision and supporting regulatory compliance requirements for clinical AI systems. The state graph enforces a maximum depth of five transitions per user turn, preventing infinite loops, and operates with deterministic edge routing that eliminates cyclic dependencies.

\subsection{Implementation}
\label{sec:implementation}

Healink is implemented as an asynchronous web service built on FastAPI. Session state persists in Redis, structured clinical data in MySQL, and vector representations for historical case retrieval in Parquet files. The state graph is implemented using LangGraph. All LLM interactions are performed through structured output schemas to enforce response formatting and prevent uncontrolled generation. Patient-identifying data fields are encrypted at rest. The system enforces timeout wrappers on all LLM calls with graceful fallback to escalation when latency thresholds are exceeded.
The system provides standardized REST API endpoints for EHR integration, with configurable schema adapters for common electronic health record formats and webhook-based escalation notifications for clinical alert systems.

\section{Conclusion}
\label{sec:conclusion}

Healink introduces a multi-agent architecture for post-discharge follow-up care that grounds every clinical assertion in verified patient records and traceable evidence. Our evaluation on 400 real-world follow-up cases across six specialties demonstrates that prescription-anchored anti-hallucination and tiered evidence retrieval improve information completeness by 17.5\% to 31.9\% over standalone foundation models. In physician-blinded review, independent clinicians rated AI-generated responses above those of their own peers, identifying a systematic completeness gap in current follow-up practice that architectural constraint enforcement can close.

The broader significance of these findings lies in two insights. First, safety in clinical AI is not a matter of better prompting but of architectural design: physical constraint enforcement downstream of generation transforms probabilistic safety aspirations into deterministic guarantees. Second, the divergence between automated LLM evaluation and physician judgment reveals that current benchmarks may systematically misrank systems optimized for practical clinical utility, calling for hybrid evaluation frameworks that combine automated screening with targeted human assessment.

The societal implications are immediate. Post-discharge follow-up is the longest phase of the patient journey, affecting hundreds of millions of patients annually, yet it remains the least structurally supported. By making high-quality follow-up support scalable and deployable with cost-effective open-source models, Healink offers a pathway to reduce physician burden while ensuring that every patient receives comprehensive guidance on medication adherence, warning signs, and recovery management. In a domain where trust is earned through transparency, the architectural commitment to accountability embodied in Healink's white-box evidence chain may prove as consequential as the performance gains it enables.

\bibliography{sn-bibliography}

\end{document}